\documentclass[twocolumn]{aastex7}

\newcommand{\heii}{\ion{He}{2}}
\newcommand{\oii}{[\ion{O}{2}]}
\newcommand{\oiii}{[\ion{O}{3}]}

\newcommand{\nii}{[\ion{N}{2}]}
\newcommand{\oiv}{[\ion{O}{4}]}
\newcommand{\ariv}{[\ion{Ar}{4}]}
\newcommand{\hb}{H$\beta$}

\newcommand{\tempoiii}{T$_{\rm e}$(\oiii)}
\newcommand{\tempoii}{T$_{\rm e}$(\oii)}
\newcommand{\tempnii}{T$_{\rm e}$(\nii)}
\newcommand{\izw}{I\,Zw\,18}
\newcommand{\hii}{H\,\textsc{ii}}

\newcommand{\denoii}{$n_{\rm e}$(\oii)}
\newcommand{\denaiv}{$n_{\rm e}$(\ariv)}

\newcommand{\Te}{$T_{\rm e}$}
\newcommand{\Ne}{$n_{\rm e}$}

\begin{document}

\title{The Interstellar Medium in \izw\ seen with JWST/MIRI: III. Spatially Resolved Three Ionization State Oxygen Abundance}

\correspondingauthor{Ryan J. Rickards Vaught}
\author[0000-0001-9719-4080]{Ryan J. Rickards Vaught}
\affiliation{Space Telescope Science Institute, 3700 San Martin Drive, Baltimore, MD 21218, USA}
\email[show]{rrickardsvaught@stsci.edu}

\author[0000-0001-9162-2371]{Leslie~K. Hunt}
\email{lesliekipphunt@gmail.com}
\affiliation{INAF -- Osservatorio Astrofisico di Arcetri, Largo E. Fermi 5, 50125 Firenze, Italy}

\author[0000-0003-4137-882X]{Alessandra Aloisi}
\email{aloisi@stsci.edu}
\affiliation{Space Telescope Science Institute, 3700 San Martin Drive, Baltimore, MD 21218, USA}

\author[0000-0002-1860-2304]{Maria.~G. Navarro~Ovando}
\email{maria.navarro@inaf.it}
\affiliation{INAF -- Osservatorio Astronomico di Roma, Via di Frascati 33, 00040 Monteporzio Catone, Italy}

\author[0000-0003-2589-762X]{Matilde Mingozzi}
\email{mmingozzi@stsci.edu}
\affiliation{Space Telescope Science Institute, 3700 San Martin Drive, Baltimore, MD 21218, USA}

\author[0000-0003-4372-2006]{Bethan James}
\email{bjames@stsci.edu}
\affiliation{Space Telescope Science Institute, 3700 San Martin Drive, Baltimore, MD 21218, USA}

\author[0000-0002-0191-4897]{Macarena G. del Valle-Espinosa}
\email{mgarciavalle@stsci.edu}
\affiliation{Space Telescope Science Institute, 3700 San Martin Drive, Baltimore, MD 21218, USA}

\author[0000-0002-4378-8534]{Karin M. Sandstrom}
\email{karin.sandstrom@gmail.com}
\affiliation{Department of Astronomy \& Astrophysics, University of California, San Diego, 9500 Gilman Dr., La Jolla, CA 92093, USA}

\author[0000-0002-8192-8091]{Angela Adamo}
\email{angela.adamo@astro.su.se}
\affiliation{Department of Astronomy, The Oskar Klein Centre, Stockholm University, AlbaNova, SE-10691 Stockholm, Sweden}

\author[0000-0002-0986-4759]{Francesca Annibali}
\email{francesca.annibali@inaf.it}
\affiliation{INAF -- Osservatorio di Astrofisica e Scienza dello Spazio, Via Gobetti 93/3, 40129 Bologna, Italy}

\author[0000-0002-5189-8004]{Daniela Calzetti}
\email{calzetti@astro.umass.edu}
\affiliation{Department of Astronomy, University of Massachusetts Amherst, 710 North Pleasant Street, Amherst, MA 01003, USA}

\author[0000-0002-0846-936X]{B.~T.~Draine}
\email{draine@astro.princeton.edu}
\affiliation{Dept.\ of Astrophysical Sciences,
  Princeton University, Princeton, NJ 08544, USA}

    \author[0000-0003-4857-8699]{Svea Hernandez}
  \email{sveash@stsci.edu}
\affiliation{AURA for ESA, Space Telescope Science Institute, 3700 San Martin Drive, Baltimore, MD 21218, USA}

\author[0000-0002-2954-8622]{Alec S.\ Hirschauer}
\email{alec.hirschauer@morgan.edu}
\affiliation{Department of Physics \& Engineering Physics, Morgan State University, 1700 East Cold Spring Lane, Baltimore, MD 21251, USA}

\author[0000-0002-0522-3743]{Margaret Meixner}
\email{margaret.meixner@jpl.nasa.gov}
\affil{Jet Propulsion Laboratory, California Institute of Technology, 4800 Oak Grove Dr., Pasadena, CA 91109, USA}

\author[0000-0001-6854-7545]{Dimitra Rigopoulou}
\email{Dimitra.Rigopoulou@physics.ox.ac.uk}
\affiliation{Department of Physics, University of Oxford, Keble Road, Oxford OX1 3RH, UK}

\author[0000-0002-0986-4759]{Monica Tosi}
\email{monica.tosi@inaf.it}
\affiliation{INAF -- Osservatorio di Astrofisica e Scienza dello Spazio, Via Gobetti 93/3, 40129 Bologna, Italy}


\begin{abstract}
We present observations of the nearby extremely metal-poor galaxy \izw\ using the Keck Cosmic Web Imager (KCWI) and the JWST Mid-InfraRed Instrument (MIRI) Integral Field Spectrographs (IFS). From optical and mid-IR oxygen emission lines, we measured direct-method abundances for three ionic states of oxygen, including O$^{3+}$/H$^+$. In contrast to previous studies of \izw\, the high spatial resolution afforded by KCWI and MIRI/MRS revealed chemical inhomogeneities on 60 pc scales in the form of metal-poor pockets and metal-enriched gas. These are located outside \izw's star-forming complexes having possibly been dispersed beyond these regions via stellar feedback effects. We found that metallicities derived using a single low-ionization density tracer, and \tempoii\ derived from a temperature relationship commonly used in high-$z$ galaxy studies, exhibited the largest scatter and underestimated the metallicity compared to those derived using multi-ion densities and estimated \tempnii. Finally, we compared O$^{3+}$/H$^+$ abundances from a theoretical ionization correction factor (ICF) against observed values and found that the oxygen ICF underestimates the O$^{3+}$/H$^+$ abundance by a factor of 2, indicating that either additional ionizing sources are needed or standard stellar population models are unable to produce the requisite ionizing flux.
\end{abstract}

\section{Introduction}
\label{sec:intro}
Galaxy chemical abundances (i.e., metallicities) are a direct consequence of their enrichment through stellar evolution channels as well as the transport of metals via gas flows \citep[e.g., inflows, outflows, and mixing;][]{Roy1995,Tenorio-Tagle1996,MacLow1999,Krumholz2018,Emerick2020}. Consequently, mapping the distribution of metallicities is key to understanding the astrophysical processes and physics that drive galaxy evolution.

Metallicity is commonly traced by the gas-phase abundance of oxygen, or other metals (e.g., sulfur, nitrogen, iron). Gas-phase abundances are estimated from ratios between ionic and hydrogen line emission both of which originate from photo-ionized gas \citep[][]{Aller1945,Piembert2017PASP}. In order to calculate the emissivities used to convert line ratios to ionic abundances (e.g., \oiii/\hb $\rightarrow$O$^{2+}$/H$^{+}$) one requires knowledge of the electron density, \Ne, and more importantly the electron temperature, \Te\ (i.e., direct-method).  Due to radially decreasing intensity, hardening of the radiation field around the ionizing source, and changes in the ions cooling the gas, ions of similar ionization potential will occupy the same volume of gas \citep[i.e., ionization zone;][]{Stasinska1980,Baldwin:2000:orion,Berg2021}. Because each ionization zone can have its own density and temperature, the accuracy of derived abundances can be sensitive to the measured/assumed properties of each zone \citep{Hagele2006,Mendez-Delgado2023Nature,Nicholls2020,Berg2021,Mingozzi2022,Hayes2025}.

Obtaining robust temperatures for each ionization zones can be challenging because the temperature-sensitive ``auroral" lines (e.g., \oiii$\lambda4363$) can be $\sim 100\times$ fainter than hydrogen emission \citep{Berg2015ApJ...806...16B}. Relationships between the temperatures of different ionization zones have been constructed using models of \hii\ regions to estimate the temperatures for ionization zones with unobserved auroral lines. However, standard modeling of \hii\ regions usually applies simplifying assumptions \citep[e.g., homogeneous gas properties;][]{Garnett1992}, and as a result may not reflect the temperature structure of physical \hii\ regions. Moreover, empirical temperature relationships, particularly those relating temperatures from \oiii\ and \oii\ auroral lines, obtained from \hii\ regions and star-forming galaxies can be biased due to temperature fluctuations, density inhomogeneities, and/or shocks \citep{Esteban2002,Tsamis2005,James2009,Binette2012,Mendez-Delgado2023Nature,RickardsVaught2024, MendezDelgado2025}. These obstacles can be amplified for high-$z$ observations where weak emission lines can be difficult to recover due to limiting flux and resolution effects. In the absence of detection of the emission lines that diagnose the physical properties of the gas, metallicity can instead be inferred from indirect `strong' line methods based on brighter ionic emission. However, the different strong-line calibrations methods disagree on the absolute metallicity by at least $\sim0.7$ dex \citep{Kewley2008} and their applicability to high-redshift and metal-poor galaxies is still a subject of ongoing investigations \citep[e.g.,][]{Croxall2013,Sanders2024,Scholte2025arXiv}.

The capabilities of JWST have greatly expanded the ability to measure direct abundances in the early universe. The rest-frame wavelengths of the standard emission lines used for direct-method chemical abundances are found in the optical between 3700~\AA$-$10,000~\AA, limiting abundances to objects whose redshifted emission is observable with optical instruments \citep[$z < 2$;][]{Tremonti2004,erb2006,Mannucci2009, Troncoso2014,Sanders2015,Sanders2016,Strom2022}. Now with the infrared coverage afforded by JWST, redshifted optical emission lines are readily observable beyond previous redshift constraints \citep[e.g.,][]{Arellano-Cordova2022,Curti2023,Laseter2024,Sanders2024, Venturi2024}. The metallicities of higher redshift galaxies are generally low \citep[$Z<0.20 Z_{\odot}$;][]{Sanders2024}, and a select few exhibit higher than expected ratios between oxygen and nitrogen abundances that challenge our understanding of chemical evolution and star formation in the early universe \citep{Bunker2023_GNz11, Cameron2023_GNz11,Arellano_Cordova2024arXiv, Marques-Chaves2024,Senchyna2024_GNz11,Massimo2024arXiv, Topping2024MNRAS,Topping2024arXiv}. Potential solutions to explain these enhancements include: massive stars within dense clusters \citep{Senchyna2024_GNz11}, rapid chemical enrichment \citep{Kobayashi2024}, Population III stars \citep{Cameron2023_GNz11, Nandal2024}, and Wolf-Rayet and Supermassive stars \citep{Charbonnel2023}. But, these solutions may be fine-tuned or involve processes not applicable to general galaxy populations. Importantly, these studies may not be using the correct densities or temperature for the ionization zones \citep{Hayes2025}.

In order to resolve the uncertainties surrounding high-$z$ metallicities, especially the abundance of triply ionized oxygen, we must turn to local systems that mimic the properties of early galaxies. Nearby Blue Compact Dwarf (BCD) galaxies are analogous to high-$z$ galaxies in stellar mass, $< 10^9 M_{\odot}$, and metallicity, $<0.05Z_{\odot}$, \citep{Schaerer2022A&A, Brinchmann2023}.  Observations of BCDs have shown that they exhibit emission from triply-ionized oxygen \citep[O$^{3+}$;][]{Hunt2010,Berg2021, CLASSY}. 
The relative contribution of O$^{3+}$ to the total oxygen abundance is uncertain for high-$z$ galaxies ($z>9.5$), because the mid-IR and UV transitions of \oiv\ can be redshifted beyond the observing capabilities of JWST. Ionization correction factors (ICFs), constructed from photoionization modeling, predict that the contribution of O$^{3+}$ is small \citep[see][]{Izotov2006}. However, these standard models fail to reproduce the observed O$^{3+}$ abundance measured in BCDs \citep{Berg2021}, and more generally fail to produce other high-ionization species \citep{Shirazi2012MNRAS.421.1043S,Kehrig2015ApJ...801L..28K,Kehrig2018, Berg2021}.

To obtain an accounting of the oxygen abundance in metal-poor BCD galaxies, we obtained JWST Mid-Infrared Instrument Medium Resolution spectrometer (MIRI/MRS) and Keck Cosmic Web Imager (KCWI) integral-field spectrographs (IFS) observations of the BCD \izw. Highly regarded as the archetype local analogue of high-redshift galaxies, \izw\ is located at a distance of 18.2 $\pm$ 1.5 Mpc \citep{Aloisi2007}, is extremely metal deficient \citep[12+log(O/H) $\sim$ 7.2;][]{Zwicky1966,Searle1972ApJ...173...25S,Skillman1993,Izotov1999ApJ...511..639I,Kehrig2016,Mingozzi2022}, has a stellar mass $\sim 10^6 - 10^7 M_{\odot}$ \citep{Fumagalli2010,Madden2014,Nanni2020}, and contains a region of vigorous star-formation between $\sim$ 0.2 M$_{\odot}$ yr$^{-1}$ \citep{Hunt2005A&A...436..837H,Annibali2013} or 0.6 M$_{\odot}$ yr$^{-1}$ based on JWST imaging \citep{Hirschauer2024,Bortolini2024}. The combined IFS observations allow for tens of pc scale resolved analysis of \izw's metallicity, including the abundance of triply-ionized oxygen. 

This is the third paper in the series that analyzes the MIRI observations of \izw. The first, Hunt et al. (hereafter Paper\,I), describes one-dimensional aperture spectra of the highly ionized lines and the newly identified 15-um continuum sources; Hunt et al. (hereafter Paper\,II) discusses the warm molecular hydrogen (i.e., H$_2$) and dust properties measured in one-dimensional spectra. Here, 
we describe the KCWI and JWST observations and reduction in Section \ref{sec:data}. The fitting and the modeling of the stellar continuum and nebular emission is detailed in Section \ref{sec:spectrum_modeling}. We describe the derivation of physical properties for the ionized gas in Section \ref{sec:derivation_of_nebular_properties}, and present our results and discuss their implications in Sections \ref{res:main}
 and \ref{disc:main}. Finally, we summarize this work in Section \ref{sec:conlusion}. Throughout this work, we assume a solar metallicity of 12+log(O/H)=8.69 \citep{Asplund2009A}.

\section{Data}
\label{sec:data}
\subsection{JWST MIRI/MRS}
\label{sec:data_miri}
\begin{figure}
    \centering
    \includegraphics[width=\linewidth]{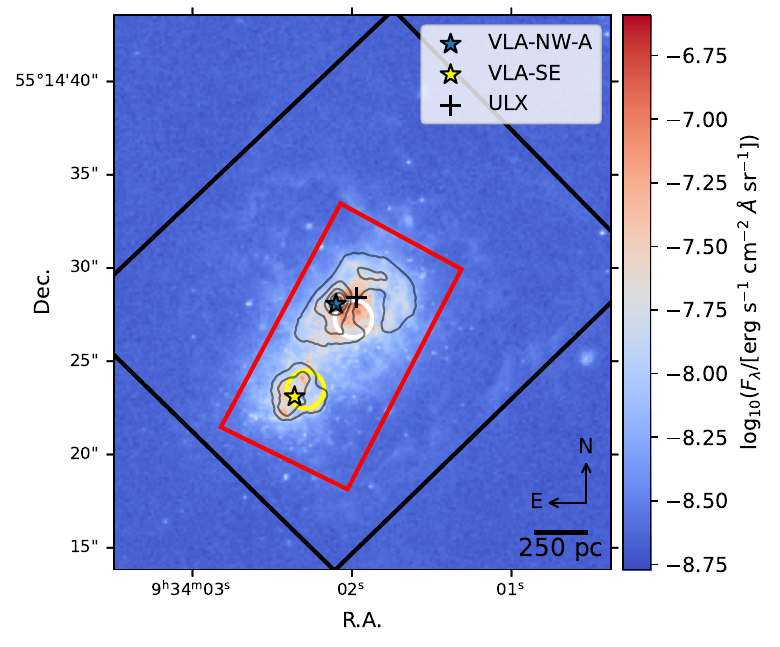}
    \caption{The Keck-KCWI and JWST-MIRI/MRS coverage of \izw. We overlay the Keck-KCWI (black) JWST-MIRI/MRS Ch.4 (red) instrument fields of view on top F606W Hubble Space Telescope imaging of  \izw\ \citep{Aloisi2007}. The yellow and white circles indicate the approximate centers of the Southeast and Northwest star forming complexes (SE and NW, Paper I). We indicate the positions of Very Large Array observations covering two \hii\ regions in the Southeast and Northwest star forming complexes \citep[VLA-SE, VLA-NW-A, see][]{Cannon2005}. We also show the position of \izw's Ultraluminous X-ray source \citep[ULX, ][]{Thuan2004ApJ...606..213T}. To help identify the emission co-spatial with the SE and NW star forming complexes, we overlay contours of H$\beta$ emission for 5 levels between 0.4-2.4 $\times 10^{-16}$ erg s$^{-1}$ cm$^{-2}$.}
    \label{fig:FoV}
\end{figure}

The \izw\ data were obtained in Cycle 2 JWST program PID 3533 (PI: Aloisi \& Co-PI Hunt). The galaxy was observed using MIRI/MRS. As shown by the red box in Figure \ref{fig:FoV}, two pointings were needed to cover the two star-forming complexes of the galaxy. A 4-pt dither was performed for each pointing. The MIRI/MRS instrument was configured to cover the mid-IR spectrum between 5$\mu$m and 28$\mu$m  which was achieved using the four channels and three grating settings (SHORT, MEDIUM, and LONG). The spectral resolving power and angular resolution of MIRI/MRS is wavelength dependent. The spectral resolving power of the MIRI/MRS instrument is well approximated by Equation 1 from \cite{Jones:2023_resolving_power} and ranges between $R\sim4000-1000$ between 5$\mu$m and 28$\mu$m. The angular resolution for each wavelength slice can be approximated using Equation 1 from \cite{Law2023}; for the line of interest we examine here, \oiv, the MIRI angular FWHM is 0\farcs96. The wavelength calibration in Channel 4C is good to 30 km s$^{-1}$. Because the galaxy completely fills the instrumental field of view, a dedicated background image was obtained for background subtraction. The final data cube has a pixel scale of 0\farcs13$\times$0\farcs13. The red box in Figure \ref{fig:FoV} shows the largest field of view (Ch. 4) covered by the MIRI/MRS observations. For additional details regarding the MIRI/MRS observations and reduction see Paper\,I.

\subsection{Keck KCWI}
\label{sec:data_kcwi}
Observations of \izw's stellar and nebular emission within the wavelength range, $3700$\AA\ - $5500$\AA, were obtained from KCWI imaging \citep{RickardsVaught2021}. The small slicer and large blue (BL) grating KCWI configuration provides a wavelength rms, 0.08\AA\ (or 5 km s$^{-1}$), spectral resolution, R$\sim$3600 (or 80 km s$^{-1}$), and pixel scale, 0\farcs15$\times$0\farcs35. The KCWI observations were reduced using the Version 1.0.1 Python implementation of the KCWI Data Extraction and Reduction Pipeline (KDRP)\footnote{\href{https://github.com/Keck-DataReductionPipelines/KCWI\_DRP}{KCWI DRP-Python}}. It was built using the Keck Data Reduction Pipeline Framework package\footnote{\href{https://github.com/Keck-DataReductionPipelines/KeckDRPFramework}{KeckDRPFramework}} and is a port of the initial IDL reduction pipeline\footnote{\href{https://github.com/Keck-DataReductionPipelines/KCWIDRP}{KCWI DRP-IDL}} \citep{2018ApJ...864...93M}.

The data consist of 4 pointings that were observed in clear conditions. Prior to mosaicking the pointings into a single image, the images of each pointing are re-projected onto a uniform 0\farcs15$\times$0\farcs15, or 13pc$\times$13pc, pixel grid. From standard star observations, the KCWI angular FWHM is 0.7\arcsec\ \citep{RickardsVaught2021}. Each image is then aligned to archival Hubble Space Telescope (HST) F606W image of \izw\ \citep{Aloisi2007}. The astrometry of the F606W image is anchored to Gaia stars within the image field of view. The alignment is performed via a pixel-grid search for offsets which optimize the correlation between the KCWI and HST image \citep[see][]{RickardsVaught2024}. After alignment, the images are co-added, resulting in a single spectral datacube of \izw. The KCWI field of view covered by the final mosaic is shown in Figure \ref{fig:FoV}. \cite{RickardsVaught2021,RickardsVaught2024} provide additional details of the KCWI observations and mosaicking.

\section{Construction of Emission-Line Maps} 
\label{sec:spectrum_modeling}

\subsection{Stellar-Continuum Modeling}
The stellar contribution to the KCWI spectra is fit, on a pixel-by-pixel basis, using {\tt LZIFU} \citep{Ho2016}. {\tt LZIFU} is an implementation of the penalized pixel fitting routine \citep[{\tt pPFX},][]{Cappellari2004} optimized for use on spectral data cubes. We fit the stellar component with SSPGeneva\footnote{For a complete description see, \href{https://home.iaa.csic.es/~rosa/research/synthesis/HRES/ESPS-HRES.html}{SSPGeneva}.} stellar templates with sub-solar metallicities \citep{GonzalezDelgado2005SSPGeneva, Martins2005}. Generally, the stellar absorption component underlying the Balmer transition in \izw\ is small with respect to the nebular emission. We refrain from modeling the stellar continuum in the MIRI mid-IR data, as the continuum is not dominated by starlight.

\subsection{Emission-Line Fitting}
\begin{figure*}
    \centering
    \includegraphics[width=0.4\linewidth]{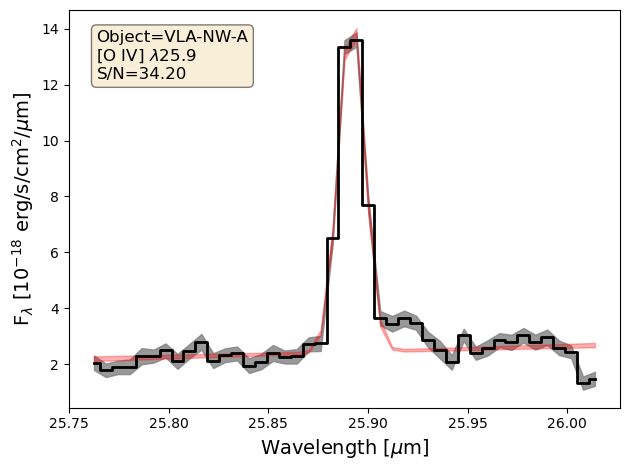}
    \includegraphics[width=0.4\linewidth]{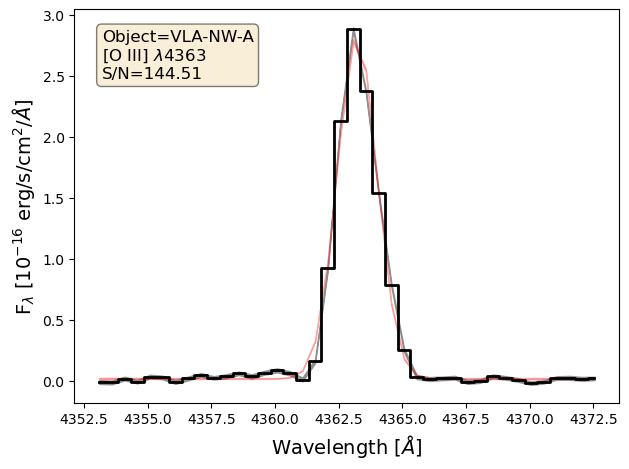}
    \includegraphics[width=0.4\linewidth]{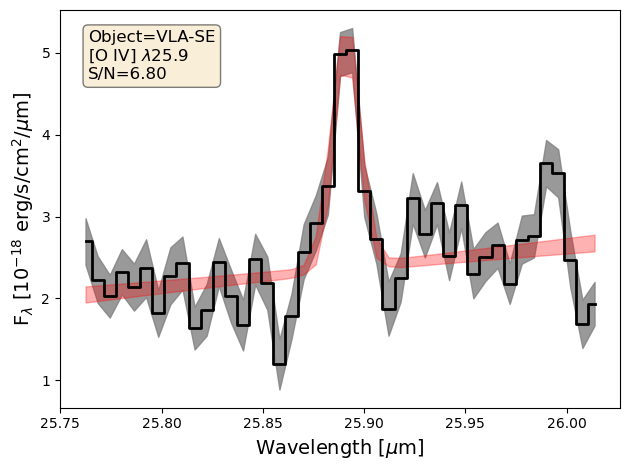}
    \includegraphics[width=0.4\linewidth]{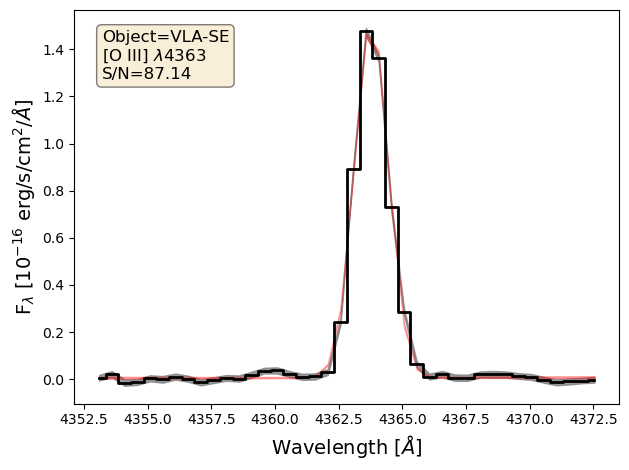}
    \caption{Gaussian plus linear continuum model fits to the \oiv$\lambda25.9\mu$m and auroral line \oiii$\lambda4363$\AA\ emission. The spectra shown in he top and bottom rows have been integrated using a $r$=0.57\arcsec\ aperture centered on the positions of VLA-NW-A and VLA-SE  (see Figure \ref{fig:FoV}). The shaded grey and red bands show the $\pm1\sigma$ of the uncertainty and fitted models. We note that the applied fitting routine is able to account for uncertainties introduced by the residual fringing around the \oiv\ emission.}
    \label{fig:fits}
\end{figure*}

The uncertainty in the continuum subtraction, and fringing present in the MIRI/MRS spectra, will dominate the uncertainty in the measured weak emission-line fluxes. After subtracting the stellar continuum, and to accurately assess the integrated fluxes and uncertainties of faint emission lines, we fit each emission line using the following framework: We determine the emission line flux and gas kinematics by fitting the emission line (both strong and weak), and residual (or underlying) continuum, in each spaxel with a Gaussian plus linear continuum model. For each spaxel, we measure the standard deviation of the continuum, $\sigma_{\text{c}}$, in two windows on each side of the line of interest. The chosen window is fixed at  10~\AA\ ($\sim$ 600 km s$^{-1}$) for the KCWI spectra, but, because of the wavelength-dependence of the MIRI-MRS spectral LSF, we set the size of the continuum window to 15$\times$ the spectral-resolution element at the location of the target line. After defining the fitting window, we then perform $N=100$ trial fits at the location of the line of interest. For each trial fit we add noise to the spectrum that is drawn from a normal distribution. The width of the normal distribution is set to standard deviation of the off-line continuum flux, $\sigma_{\text{c}}$. After the $N$ trials are complete, we take the average of the measured flux distribution as the measured line flux, $F_{\lambda}$. The uncertainty on the line flux is the standard deviation of the flux distribution, $\sigma_{F}$. We show in Figure \ref{fig:fits} an example of low and high S/N fits to mid-IR \oiv\ and optical \oiii\ auroral emission.

\begin{figure*}[t]
    \centering
    \includegraphics[scale=0.6]{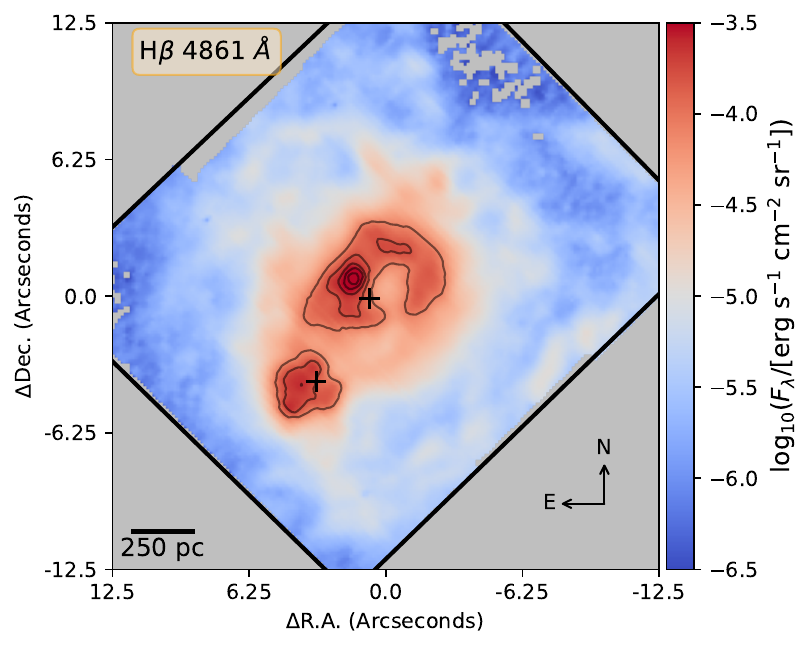}
    \includegraphics[scale=0.6]{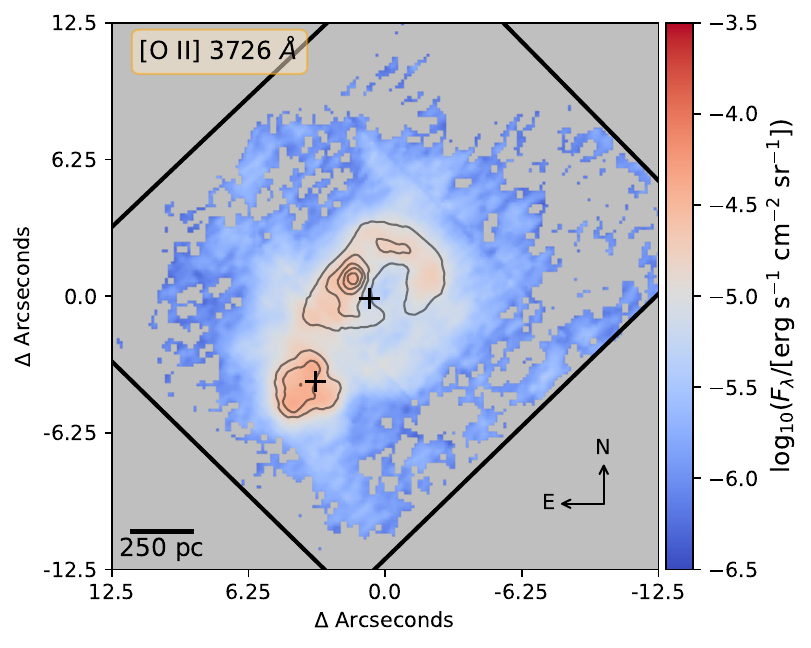}
    \includegraphics[scale=0.6]{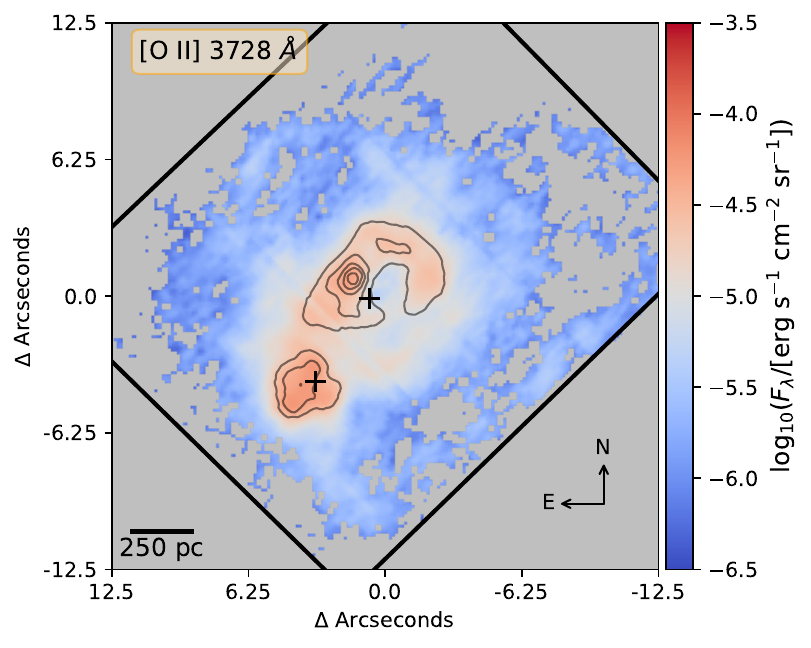}
    \includegraphics[scale=0.6]{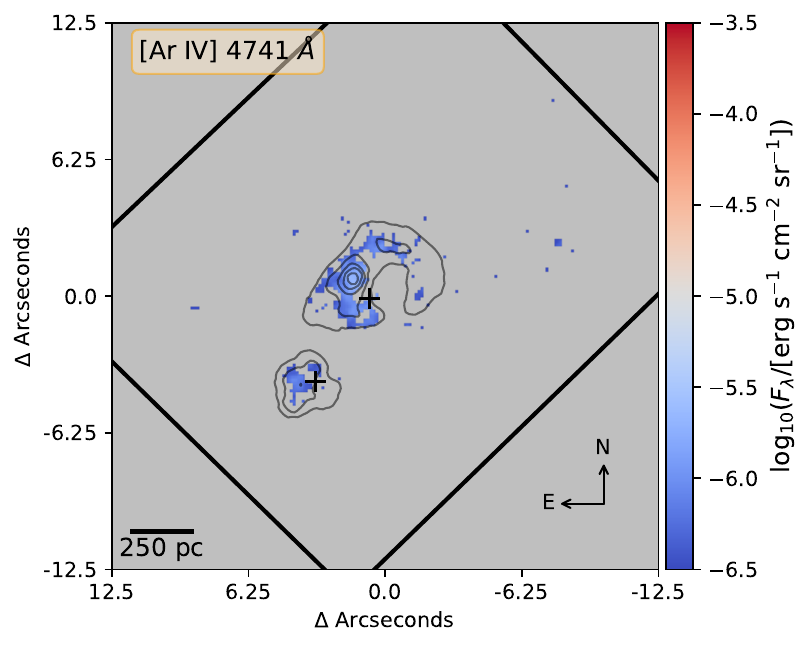}
    \includegraphics[scale=0.6]{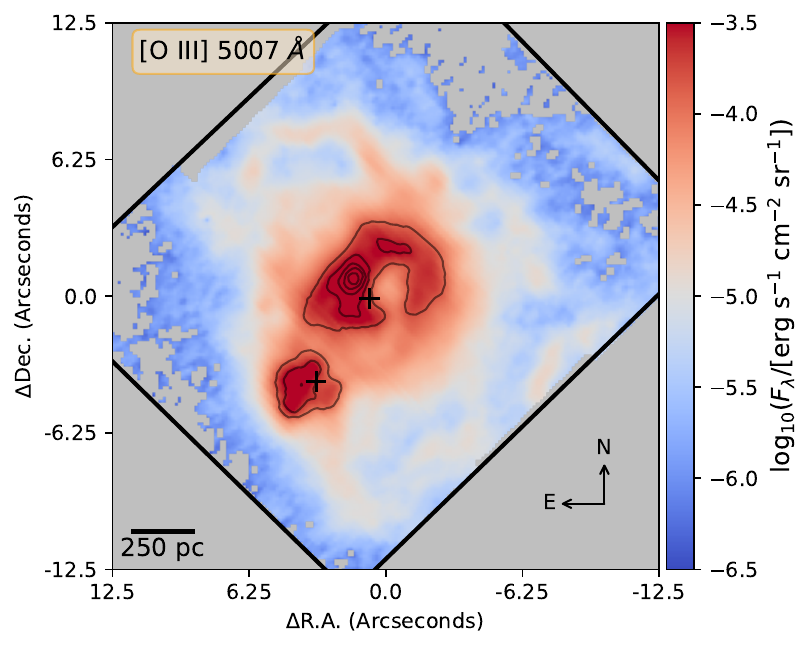}
    \includegraphics[scale=0.6]{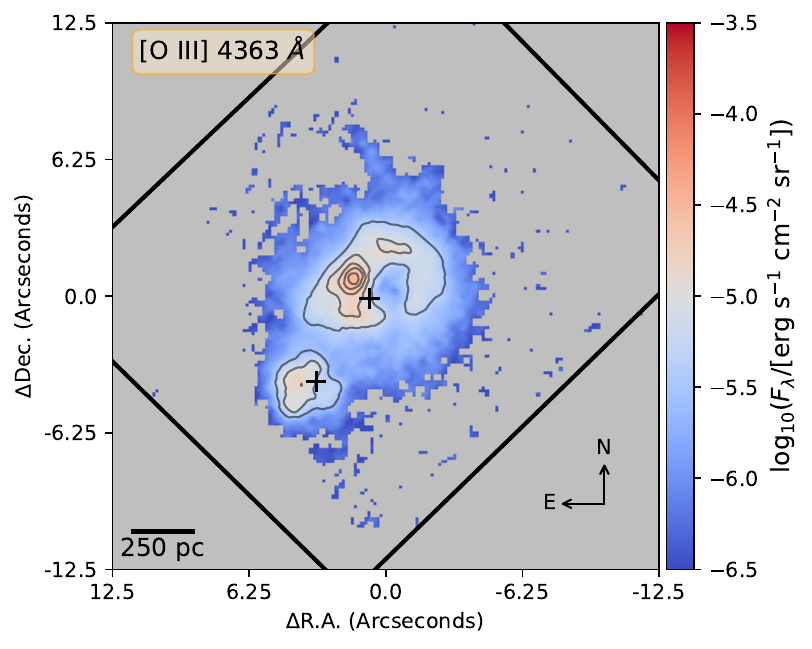}
    \caption{Selection of optical emission-line maps observed in KCWI. Only pixels with S/N $>$ 3 are shown. The emission traces the ionized gas around the Northwest and Southeast stellar OB associations (see Figure \ref{fig:FoV}), as well as filamentary structure surrounding them. The R.A. and Dec. offset are centered on the coordinates 09$^{\rm h}$~34$^{\rm m}_.$0~01$^{\rm s}_.$92, 55$^{\circ}$~14\arcmin ~27.4\arcsec. To help identify the emission co-spatial with the SE and NW star forming complexes (black-crosses), we overlay contours of H$\beta$ emission for 5 levels between 0.4-2.4 $\times 10^{-16}$ erg s$^{-1}$ cm$^{-2}$. We also overlay the KCWI field of view (black-box).}
    \label{fig:optical_line_maps}
\end{figure*}

\begin{figure*}[t]
    \centering
    \includegraphics[scale=0.6]{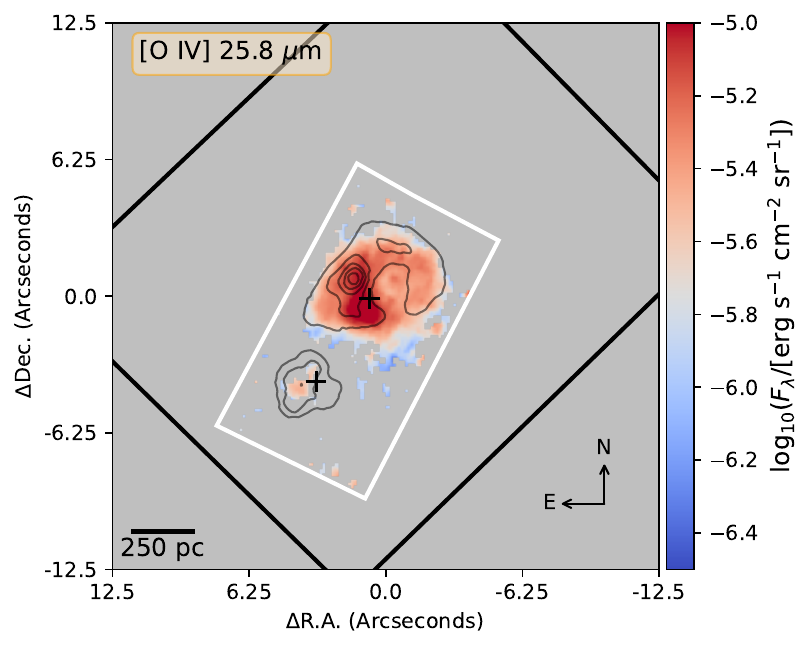}
     \includegraphics[scale=0.6]{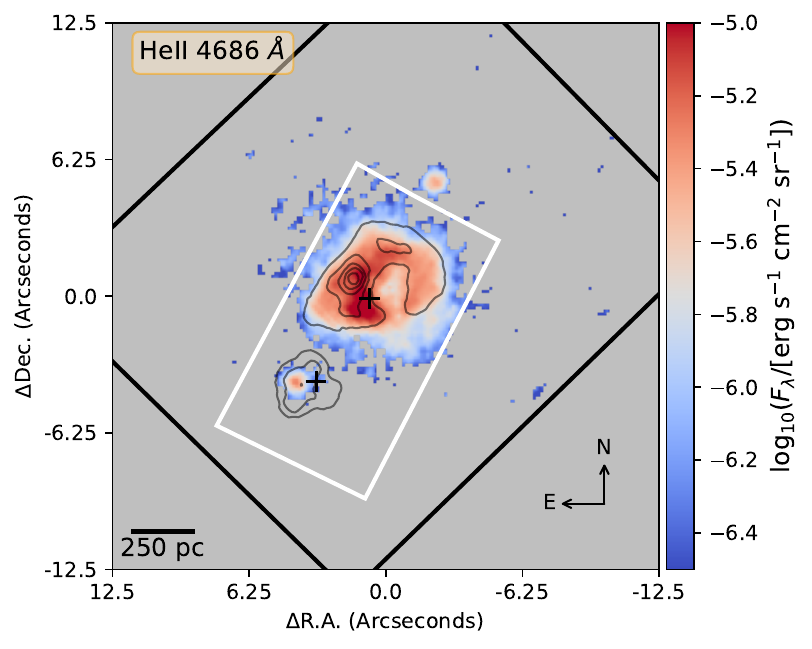}
     \caption{Emission-line maps of the high-ionization, E$>$54 eV, transitions \oiv$\lambda25.89\mu$m and \heii$\lambda4686$\AA\ as observed in MIRI/MRS and KCWI.  Only pixels with S/N $>$ 3 are shown. The R.A. and Dec. offset are centered on the coordinates 09$^{\rm h}$~34$^{\rm m}_.$0~01$^{\rm s}_.$92, 55$^{\circ}$~14\arcmin ~27.4\arcsec. To help identify the emission co-spatial with the SE and NW star forming complexes (black-crosses), we overlay contours of H$\beta$ emission for 5 levels between 0.4-2.4 $\times 10^{-16}$ erg s$^{-1}$ cm$^{-2}$. We also overlay the KCWI field of view (black-box) and MIRI (white-box) fields of view.}
     \label{fig:very_high_maps}
\end{figure*}

We also fit the strong-line emission (e.g., H$\beta$, \oiii$\lambda5007$) using the above framework. However, we note  here that the uncertainty in the flux of strong lines is dominated by uncertainties in the absolute flux calibration rather than the continuum subtraction. When reporting absolute fluxes of strong lines, we add in quadrature a 5\% flux uncertainty with the fitting error to take into account the reported uncertainty in the MIRI/MRS Channel 4 flux calibration \citep{Law2025_Miri_flux_cal}. Additionally, and assuming a systemic velocity of 751 km s$^{-1}$  \citep[$z$=0.002505;][]{Thuan1999A&AS}, we construct velocity maps from the average fitted central wavelength. We show in Figure \ref{fig:optical_line_maps} and Figure \ref{fig:very_high_maps} the resulting emission-line maps for low to high-ionization and very high-ionization species.

The blended emission lines [Ar IV]$\lambda4711$/He I $\lambda4715$ are marginally spectrally resolved. To fit their emission we perform a two Gaussian plus linear continuum fit. In the particular case of [Ar IV]$\lambda4711$/He I $\lambda4715$ blend, to accurately remove the He I $\lambda4715$ contamination from [Ar IV]$\lambda4711$, we first fit He I $\lambda4471$ and assume a theoretical ratio He I $\lambda4471$/$\lambda4715=0.150$ ($n_{\rm e}=10^2$ cm$^{-3}$, $T_{\rm e}=20,000$ K), to fix the amplitude, velocity and width of the Gaussian fit to He I $\lambda4715$. Discussed in detail in Section \ref{meth:dust}, the dust extinction in \izw\ is low thus for deblending purposes the observed ratios between different helium recombination lines should be close to their theoretical value. We show example fits to the blended [Ar IV]$\lambda4711$/He I $\lambda4715$ emission, and map of the de-blended [Ar IV]$\lambda4711$ emission in Appendix \ref{Appendix:Blended_Argon} Figure \ref{fig:blended_ar_fit}.

\section{Nebular Properties}
\label{sec:derivation_of_nebular_properties}
The optical and mid-IR spectra of \izw\ contains emission from metal and hydrogen transitions that can be used to infer the physical properties of the ionized gas. We use \texttt{PyNeb} \citep{Luridiana2015A&A...573A..42L} on a pixel-by-pixel basis to convert the measured line emission to the following physical properties: V-band dust extinction, electron density, electron temperature, and metallicity. The uncertainty on each physical parameter is propagated via sampling the uncertainty of the line emission. We summarize in Table \ref{tab:ionization_potentials} the ionization potentials and transitions of the ions used to derive the physical properties of the nebular gas.

\subsection{Dust Extinction}
\label{meth:dust}
The details describing the calculation of \izw's intrinsic V-band extinction, $A_V$, via optical and mid-IR \ion{H}{1} emission lines are described in Paper I. Correcting for Milky Way extinction, the extinction intrinsic to \izw\ is largely consistent with $A_V=0$ and is in agreement with previous determinations of $A_V$ in \izw\ \citep[see][Paper I]{Cannon2002ApJ...565..931C,Fisher2014Natur.505..186F}. Consequently, we do not dust-correct the optical to mid-IR line emission.

\subsection{Electron Density}
\label{meth:electron_dens}
\begin{figure}
    \centering
    \includegraphics[width=\linewidth]{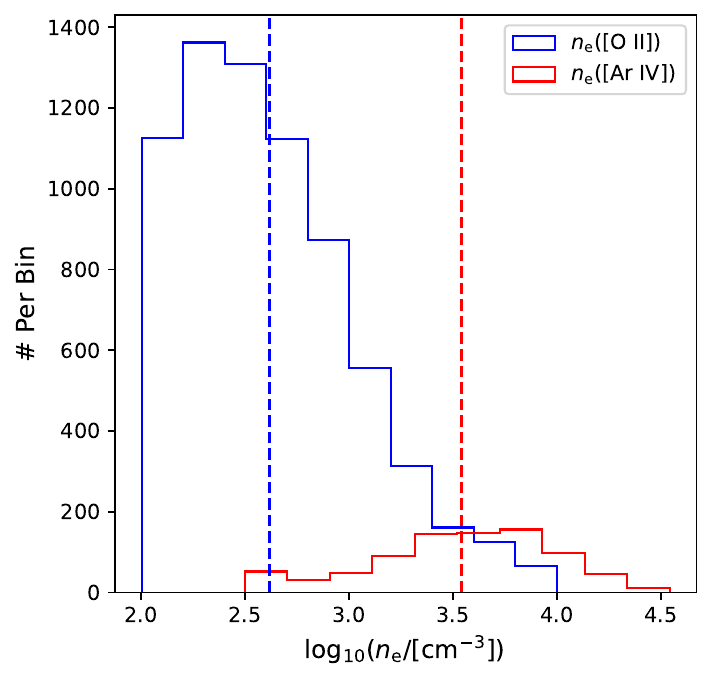}
    \includegraphics[width=\linewidth]{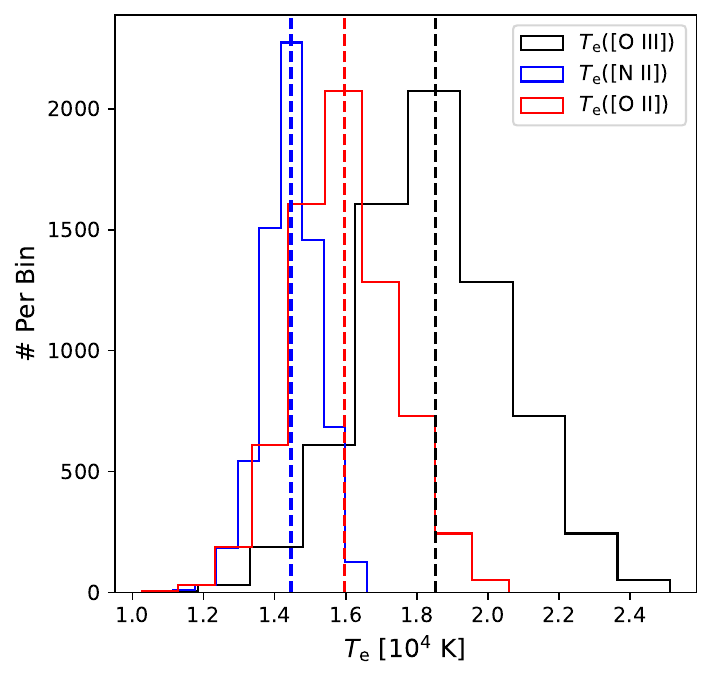}
    \caption{Densities and temperatures assumed in derivation of oxygen abundances. Top: Distribution of low- and high-ionization electron density. The blue/red histogram and dashed vertical line are the distribution and mean low/high-ionization electron density traced by the \oii\ and \ariv\ doublets. Not shown are spaxels whose densities are above or below the high- and low-density limits. Bottom: Distribution of low- and high-ionization zone electron temperature. The black histogram and dashed vertical line is the \tempoiii\ derived from the \oiii$\lambda4363$ auroral line. We also show the distribution and averages for the low-ionization zone \nii\ and \oii\ temperatures inferred from the temperature relationships of \cite{Arellano-Cordova2020} and \cite{Campbell1986}. Supplement figures showing the spatial distribution of the measured densities and temperature are presented in Appendix \ref{appendix:den_temp_maps} Figures \ref{fig:density_maps} and \ref{fig:temp_map}.}
    \label{fig:neb_prop_hist}
\end{figure}

We obtain estimates of the electron density, $n_e$, for low- and high-ionization gas using the optical density diagnostics [O II]$\lambda3729/\lambda3726$ and [Ar IV]$\lambda4740/\lambda4711$. For mid-IR density diagnostic (e.g., [\ion{Fe}{2}]$\lambda5.3/\lambda25.9$ and [\ion{Ar}{3}]$\lambda8.9/\lambda21.8$) only one line of each doublet is detected. We show the histogram of the low- and high-ionization gas densities in the top panel of Figure \ref{fig:neb_prop_hist}.

\begin{deluxetable}{ccc}
\caption{Ionization potential energies, and transitions, of the ions used in the derivation of nebular properties.}
\tablehead{ \colhead{Ion} & \colhead{Ionization Potential\tablenotemark{a}} & \colhead{Obs. Transition}
\\ \colhead{} & \colhead{(eV)} & \colhead{(\AA)}}
\startdata
H$^+$ & 13.59 & 4861 \\
O$^{+}$  & 13.61& 3726, 3728 \\
He$^+$ & 24.58& 4471 \\
O$^{2+}$ & 35.12 & 4363, 4959, 5007 \\
Ar$^{3+}$ & 45.81& 4711, 4741 \\
He$^{2+}$ & 54.42& 4686 \\
O$^{3+}$ & 54.93& 258900  \\
\enddata
\tablenotetext{a}{Ionization potential energies are obtained from Appendix D in \cite{Draine2011piim.book.....D}}
\label{tab:ionization_potentials}
\end{deluxetable}

The [O II] diagnostic traces the density of low-ionization gas, $n_{\rm e}$(\oii) and is sensitive to densities between 100 cm$^{-3}$ and 10,000 cm$^{-3}$. Of the pixels with S/N $>$ 3 in both [O II]$\lambda3729$,$\lambda3726$ emission, we find that 53\% of the total spaxels exhibit densities lower than the low-density limit, $n_{\rm e} < 100 $ cm$^{-3}$. These pixels are coincident with \izw's diffuse ionized gas around the star forming complexes. Although \texttt{PyNeb} will return values between 1 and 100 cm$^{-3}$ for these pixels,  we fix their densities to a value of $n_{\rm e} = 100 $ cm$^{-3}$ since any value below this limit is uncertain. The spaxels with measurable density have values that range between $n_{\rm e} =100 - 10,000$ cm$^{-3}$, with mean value of $\sim 400$ cm$^{-3}$ and are located within the cavity of the H$\beta$ shell surrounding the NW star forming complex (see H$\beta$ contour in Figure \ref{fig:FoV}).  

The high-ionization gas density is estimated from the [Ar IV]$\lambda4740$,$\lambda4711$ ratio, $n_{\rm e}$(\ariv). The number of spaxels with measurable emission from both lines is considerably smaller than those with \oii$\lambda3729$,$\lambda3726$ emission and largely located in spaxels with both He II and \oiv\ emission (see Figure \ref{fig:very_high_maps}). For spaxels with measured densities, the value ranges between $n_{\rm e} =300 - 31,000$ cm$^{-3}$, with mean value of $\sim 3000$ cm$^{-3}$. Maps of the measured densities can be found in Appendix \ref{appendix:den_temp_maps} Figure \ref{fig:density_maps}.

\subsection{Electron Temperature}
\label{meth:electron_temp}
We measure the electron temperature of the high-ionization gas using the temperature-sensitive line ratio \oiii$\lambda4363$/$\lambda5007$. Shown in the right bottom panel of Figure \ref{fig:optical_line_maps}, we detect \oiii$\lambda4363$ emission in a significant number of spaxels covering both the SE and NW star forming complexes as well as regions with extended nebular emission. An estimate of the electron density is required to convert the \oiii$\lambda4363$/$\lambda5007$ ratio to a temperature. Although the \oiii$\lambda4363$/$\lambda5007$ ratio is typically insensitive to density in low \Ne\ environments, the ratio gains a small secondary dependence in environments with high electron densities ($n_{\rm e} > 10^3\ \rm{cm}^{-3}$). When possible, we use the high-ionization zone density from $n_{\rm e}$([Ar IV]) because of similar ionization potentials to produce O$^{2+}$ and Ar$^{3+}$. However, for spaxels without a measurement of the high-ionization zone density, we substitute the low-ionization zone density \denoii. A histogram of \tempoiii\ is shown in the bottom panel of Figure \ref{fig:neb_prop_hist}. We measure an average temperature of \tempoiii=$1.8517 \pm 0.20\ \times 10^4$ K, a value similar within errors to \tempoiii$\sim2.0\ \times10^4$ K measured by \cite{Izotov1999ApJ...511..639I} and \cite{Kehrig2016}. A map of the \tempoiii\ measured from oxygen an be found in Appendix \ref{appendix:den_temp_maps} Figure \ref{fig:temp_map}.

The auroral lines \oii$\lambda\lambda7320,7330$ that diagnose \tempoii, a tracer of the low-ionization temperature, are outside the wavelength range of MIRI and KCWI. From observations of spiral galaxy \hii\ regions, many linear relationships have been constructed to estimate the temperature of the gas containing singly-ionized oxygen, \tempoii, from \tempoiii\ \citep[e.g.,][]{Campbell1986,Garnett1992,Berg2020,RickardsVaught2024,Brazzini2024}. However, analysis of very deep \hii\ region spectra instead show that T$_{\rm e}$([O II]) can be biased by unresolved density inhomogeneities, and that T$_{\rm e}$([N II]) is a more robust tracer of the low-ionization gas temperature \citep{Mendez-Delgado2023Nature,Mendez-Delgado2023}. For these reasons, we assume the low-ionization zone temperature is traced by the nitrogen (i.e., \tempoii=\tempnii), and calculate \tempnii\ using the \tempnii$-$\tempoiii\ relationship from \cite{Arellano-Cordova2020}. This temperature relationship, see Eq. \ref{eqt:tenii}, depends on both the \tempoiii,in units of $10^4$ K, and the degree of ionization, $P=$\oiii/(\oii+\oiii),

\begin{eqnarray}\label{eqt:tenii}
\frac{1}{T_e(\text{[N II]})}  = 
\begin{array}{cc}
\frac{0.54}{T_e(\text{[O III]})}+0.52\ & \text{,}\ P<0.5 \\ 
\frac{0.61}{T_e(\text{[O III]})}+0.36\ & \text{,}\ P\geq 0.5
\end{array}
[\frac{1}{10^4 \rm K}].
\end{eqnarray}
Additionally, to investigate how the above assumption affects the resulting metallicities, we also derive two-zone metallicities using a single density diagnostic, and \tempoii\ calculated using the temperature relationship, Eq. \ref{eqt:teoii}, from \cite{Campbell1986}. This temperature relationship is one of the commonly used relationships used in high-$z$ studies \citep[e.g.,][]{Sanders2024,Scholte2025arXiv}.

\begin{equation}
\label{eqt:teoii}
T_e(\text{[O II]})=0.7\times T_e(\text{[O III]})+0.30\ [10^4 \rm K].
\end{equation}

 We display a comparison of the three assumed temperatures in the bottom panel of Figure \ref{fig:neb_prop_hist}. The T$_{\rm e}$([N II]) temperatures are on average lower than the T$_{\rm e}$(\oii)), and exhibit a more narrow range of temperatures. As found in previous studies of \hii\ regions \citep[see,][]{Berg2020}, the \tempnii\ and \tempoii\ temperatures are cooler than \tempoiii.

\subsection{Oxygen Abundance}
\label{meth:oxygen_abundance}
We derive the oxygen abundance in all spaxels with significant detection of [O III]$\lambda4363$ emission (S/N $>3$). Due to the MIRI/MRS coverage of the mid-IR spectrum, we are in a unique position to compute the abundances O$^+$,  O$^{2+}$, O$^{3+}$ (i.e., three-state oxygen abundance);
\begin{equation}
    \frac{\rm O}{\rm H}= \frac{\rm O^{+}}{\rm H^{+}}+\frac{\rm O^{2+}}{\rm H^{+}}+\frac{\rm O^{3+}}{\rm H^{+}},
\end{equation}
where the ionic abundances of O$^+$,  O$^{2+}$, O$^{3+}$, and H$^{+}$ are measured from \oii, \oiii, \oiv, and \hb\ emission. 

The pixel scale of MIRI/MRS and KCWI are 0.13\arcsec\ and 0.14\arcsec\ respectively. To compare the \oiv$\lambda25$ emission with H$\beta$ we reproject the [O IV] map onto KCWI pixel grid using the \texttt{astropy reproject.interp} function. In deriving O$^{3+}$/H$^{+}$ and O$^{2+}$/H$^{+}$, we use high-ionization zone density $n_{\rm e}$([\ion{Ar}{4}]) or $n_{\rm e}$([\ion{O}{2}]) when the argon diagnostic is unavailable, and T$_{\rm e}$([O III]). We note that the above density replacement implicitly assumes that the high-ionization zone gas in these spaxels have the same density as the low-ionization zone, and may potentially lead to an underestimate of the O$^{3+}$/H$^{+}$ abundance in these spaxels. For O$^{+}$/H$^{+}$, we use $n_{\rm e}$([\ion{O}{2}]) and T$_{\rm e}$([N II]). The error corresponding to each pixel's metallicity is determined by sampling the uncertainty in all the requisite emission lines. The resulting three-state 12+log(O/H) (i.e., metallicity) map for \izw\ is presented in the left panel of Figure \ref{fig:metal_maps}.

To investigate how different assumptions of the ionized gas electron density and temperature impact the metallicity, we also compare the three-state metallicities to the following metallicities:
\begin{itemize}
    \item `Two-State Metallicity'- We assume no contribution from triply-ionized oxygen (i.e., O$^{3+}$/H$^{+}$=0), and make the same assumptions for density and temperatures as for three-state metallicities.

    \item `Simple Metallicity'- We assume no contribution from triply-ionized oxygen, and  assume a single density, $n_{\rm e}$([\ion{O}{2}]), using electron temperatures, T$_{\rm e}$([O II]) from Eq.\ref{eqt:teoii}, and T$_{\rm e}$([O III]), for O$^{+}$/H$^{+}$ and O$^{2+}$/H$^{+}$ respectively.
\end{itemize}

The comparisons between the three different metallicities are discussed in Section \ref{sec:3_vs_2_zone_metallicity}.

\subsection{Helium Abundance}
\label{sec:helium_abundances}
\begin{figure*}
    \centering
    \includegraphics[width=0.46\linewidth]{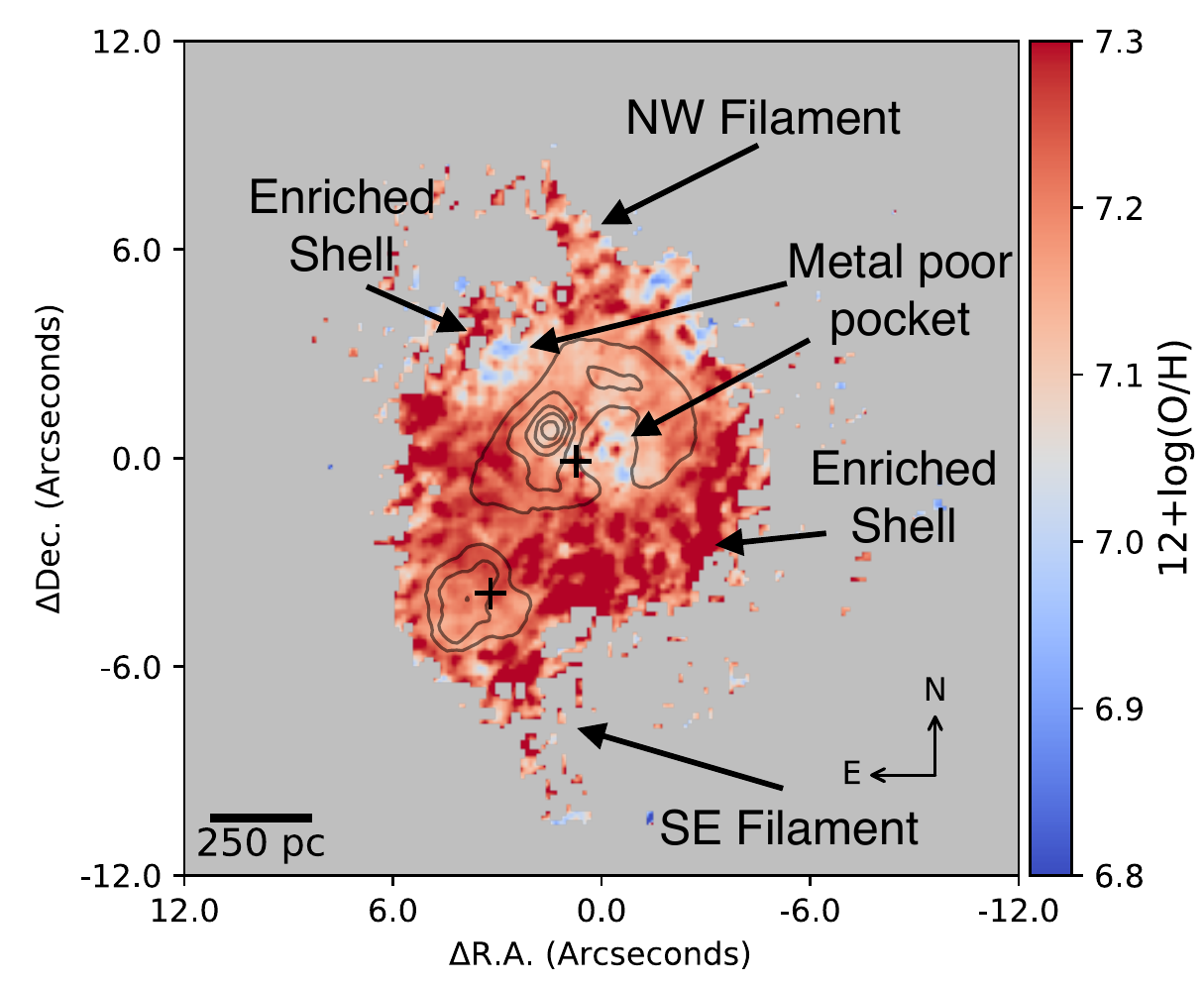} 
    \includegraphics[width=0.465\linewidth]{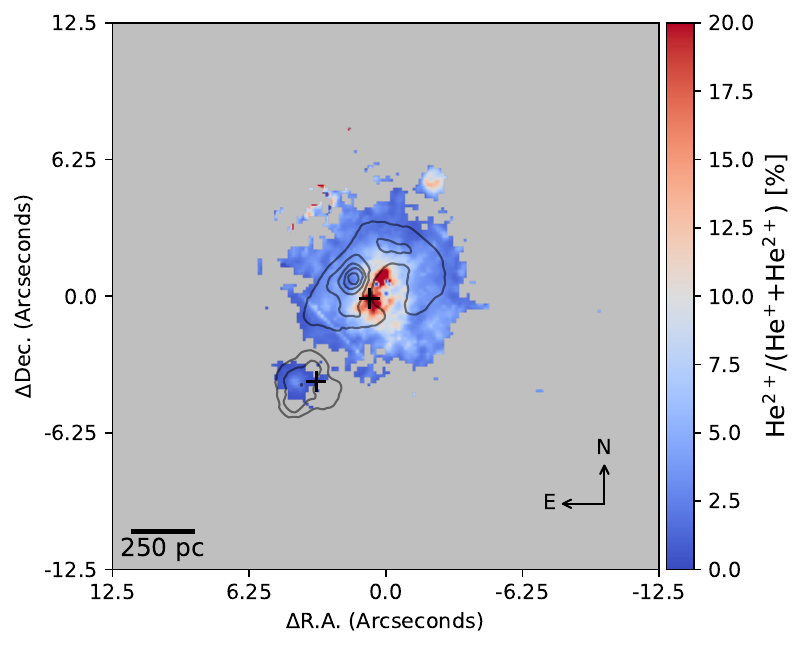}
    \caption{Left: The spatial distribution of \izw's three-state oxygen abundance or metallicity. We annotate notable features across the metallicity map including metal-poor pockets, as well as enriched shells or filaments. Right: The fraction doubly-ionized to total helium abundance. The R.A. and Dec. offset are centered on the coordinates 09$^{\rm h}$~34$^{\rm m}_.$0~01$^{\rm s}_.$92, 55$^{\circ}$~14\arcmin ~27.4\arcsec. To help identify the emission co-spatial with the SE and NW star forming complexes (black-crosses), we overlay contours of H$\beta$ emission for 5 levels between 0.4-2.4 $\times 10^{-16}$ erg s$^{-1}$ cm$^{-2}$.}
    \label{fig:metal_maps}
\end{figure*}

Doubly-ionized helium and triply-ionized oxygen both require ionizing photons with energy $\sim$ 54eV. Due to their similar ionizing potentials, the fractional amount of doubly-ionized to the total helium abundance is used to infer the abundance of triply-ionized oxygen abundance when \oiv\ emission is not observed \citep[e.g.,][]{Izotov2006}. In order to compare the observed O$^{3+}$ abundance to ionization correction factors (see Section \ref{sec:icfs}), we calculate the helium abundance using the optical transitions He I $\lambda 4471$ and He II $\lambda 4686$. For the density and temperature, we assume that \denaiv\ and \tempoiii\ describe the properties of the doubly ionized helium, but in spaxels without \denaiv, we use \denoii. For singly-ionized helium, whose ionization potential is higher than that of singly-ionized oxygen (see Table \ref{tab:ionization_potentials}), we assume \denoii\ and \tempoiii. As shown in right panel of Figure \ref{fig:metal_maps}, we construct using the  He$^{+}$/H$^{+}$ and He$^{2+}$/H$^{+}$ a map of the fractional helium abundance He$^{2+}$/(H$^{+}$+He$^{2+}$). The fractional helium abundance varies across the galaxy, peaking inside the cavity of the ionized gas shell traced by H$\beta$ emission, as well as in a region northwest of the NW star forming complex \citep[see region HT10 in][]{Hunter1995ApJ...452..238H,RickardsVaught2021}. 

To gauge the accuracy of the derived helium abundances,
we calculate the helium mass fraction, using Eq. 20 from \cite{Pagel1992MNRAS.255..325P}, within $r=1$\arcsec\ apertures centered on the SE and NW star forming complexes, see Figure \ref{fig:FoV}. Using either set of physical assumptions, we find for the star forming complexes a helium mass fraction of $Y({\rm NW})=0.16\pm 0.03$, and $Y({\rm SE})=0.23\pm 0.02$. From single-slit observations of \izw\  \cite{Izotov1998ApJ...497..227I} derived $Y({\rm NW})$ and $Y({\rm SE})$ using \ion{He}{2}$\lambda46864$ and emission from three \ion{He}{1} transition (e.g., $\lambda4471, \lambda5876, {\rm and}\ \lambda6678$). Depending on the choice of \ion{He}{1} transition (e.g., $\lambda4471, \lambda5876, {\rm and}\ \lambda6678$) the authors found that $Y({\rm NW})$ varied between 0.18 $-$ 0.23 dex, with weighted mean of 0.21 dex, and $Y({\rm SE})$ remained stable around 0.23 dex. They attribute the variation of $Y({\rm NW})$ to enhanced underlying stellar absorption. Absent spectral coverage of \ion{He}{1}$\lambda5876$ and  \ion{He}{1}$\lambda6678$, we are unable to determine the potential impact of stellar absorption on $Y({\rm NW})$. 

Because the \cite{Izotov1998ApJ...497..227I} observations consist of a single 2\arcsec$\times$300\arcsec\ slit across the galaxy, we test how the use of $r=1$\arcsec\ aperture may be impacting the NW comparison. We test this by measuring $Y({\rm NW})$ using all of \ion{He}{2} emitting spaxels within the NW star forming complex. Using these spaxels, we obtain a helium mass fraction, $Y({\rm NW})=0.21\pm0.03$, this value is consistent with the weighted mean of 0.21 dex \cite{Izotov1998ApJ...497..227I}. 

Our measured $Y({\rm NW})$ and $Y({\rm SE})$ are consistent with the previous \izw\ measurements reported by \cite{Izotov1998ApJ...497..227I}. With respect to the primordial helium abundance $Y_{p}\approx0.24$ \citep{Valerdi2021MNRAS.505.3624V,Weller2025_Yp} the value  $Y({\rm SE})=0.23\pm 0.02$ dex is in good agreement. However, the value across the \heii\ emitting spaxels, $Y({\rm NW})=0.21\pm0.03$, is lower, potentially due to stellar absorption, but consistent within errors.

\section{Results}
\label{res:main}
\subsection{Nebular Emission}
\label{res:neb_emission}
We present maps of optical emission from low to high-ionization lines in Figure \ref{fig:optical_line_maps}, as well as the emission-line maps for the very-high ionization species He$^{2+}$ and O$^{3+}$ in Figure \ref{fig:very_high_maps}. The ionized gas around the NW and SE OB complexes extends outward and well into the galaxy's extended envelope \citep{Davidson1989,Hunter1995ApJ...452..238H,Dufour1996ASPC...98..358D,Izotov2001a,Papaderos2002}. As shown by the H$\beta$ emission contours, the low- and high-ionization emission traces a shell surrounding the NW star forming complexes, and emission arising out of both star forming complexes trace a filamentary structure extending outward $>$500 pc. The existence of highly ionized species (e.g. O$^{2+}$), $>$500 pc away from the star forming complexes means either ionizing photons are able to freely travel such distances or that there are additional sources of ionization (e.g., shocks) which have created the highly ionized species. An exploration of potential sources powering ionization at these distances will presented in Paper V (Rickards Vaught et al., in prep).

In contrast to the maps of ions with ionization potentials less than $54$ eV shown in Figure \ref{fig:optical_line_maps}, the \heii\ and \oiv\ emission are not as extended. Around the NW star forming complex the very high-ionization emission extends $\sim$300 pc. On the other hand, the very high-ionization emission around the SE star forming complex is contained within a region with a $\sim 60$ pc diameter. Although it may appear that the \oiv\ emission is slightly less extended than the \heii\ emission, this is likely due to the differences in the flux sensitivity of the two instruments. From the estimated uncertainty maps, we find the 3$\sigma$ limiting flux to be $\sim6.6\times10^{-18}$ erg s$^{-1}$ cm$^{-2}$ arcsec$^{-2}$ and $\sim5.3\times10^{-17}$ erg s$^{-1}$ cm$^{-2}$ arcsec$^{-2}$ for \heii\ and \oiv\ respectively. For additional comparison, the limiting flux of \oiii$\lambda5007$ is  $\sim2.6\times10^{-18}$ erg s$^{-1}$ cm$^{-2}$ arcsec$^{-2}$, almost an order of magnitude deeper than \oiv. 

Previous Spitzer observations were the first to reveal the presence of mid-IR \oiv\ emission in \izw\   \citep{Wu:2007_SpitzerIZw18}, but, it is now spatially resolved for the first time (Hunt et al. in prep). \oiv\ is found in both star-forming complexes, and is co-spatial with \heii\ emission \citep{Kehrig2015ApJ...801L..28K,RickardsVaught2021}. The \oiv\ flux comparison between Spitzer IRS and MIRI is good to 70-80\% when compared with \cite{Wu:2007_SpitzerIZw18} and \cite{Hao2009}, see Paper\,IV. Of the three distinct regions with \heii\ emission, only two (SE and NW complexes) are within the MIRI/MRS field of view (see Figure \ref{fig:very_high_maps}). But, due to the similar ionization potentials of O$^{3+}$ and He$^{2+}$, and the high abundance of doubly-ionized helium (see Figure \ref{fig:metal_maps}), we anticipate that \oiv\ is located in the region northwest of the NW star forming complex as well. 

The relative distributions of very high-ionization emission may potentially reflect the different stellar populations hosted within each complex. The NW star forming complex is largely dominated by \izw's youngest stellar population \citep[age $< 30$ Myr,][]{Aloisi1999,Hirschauer2024,Bortolini2024}, and may tentatively host massive and/or Wolf-Rayet stars \citep{Hunter1995ApJ...452..238H,Legrand1998,1997ApJ...487L..37I,Brown2002ApJ...579L..75B}, though an older population may be unobserved due to the brightness and crowding of the younger population \citep{Bortolini2024}. The SE star forming complex is a mixture of young and older populations \citep[$0.03<$ age $< 1$ Gyr;][]{Annibali2013,Hirschauer2024,Bortolini2024}. Despite the young population of stars in both complexes, their contribution to the production of very high-ionization species is uncertain, and despite exhaustive searches for objects capable of producing the requisite high ionization photons \citep[e.g., Wolf-Rayet, shocks, ULX;][]{Kehrig2015ApJ...801L..28K, RickardsVaught2021}, the identity of the source(s) remain elusive. An investigation into the identity of the potential ionization sources using diagnostic optical and the mid-IR emission line ratios will be presented in a forthcoming analysis (Rickards Vaught et. al in prep).

\subsection{The Chemical Abundance of \izw}
Enabled by the large spectral coverage afforded by both MIRI/MRS and KCWI, we observe emission from three ionization stages of oxygen, which we use to derive the 3-zone oxygen abundance. In the following sections, we discuss the statistics of the metallicity field across the galaxy and its OB associations, as well as its structure with respect to the kinematics of the line emission, and report the presence of chemical inhomogeneities within \izw.

\subsubsection{Global and region chemical abundances}
\label{res:global_metals}
We present in Table \ref{tab:metal_table} measurements of the average global, and regional specific, for the complete set of derived metallicities. We measure the global average three-state metallicity for \izw\ to be, 12+log(O/H)=7.19 with a dispersion of 0.13 dex around the mean. The global average for the two-state and simple metallicities are both 12+log(O/H)=7.18. The dispersion for the simple metallicities is 0.30 dex, and exceeds the 0.20 dex dispersion of the two-state metallicity. 
The above average metallicities are $\sim$0.08 and 0.07 dex larger than a previous IFS direct-method estimate of metallicity 12+log(O/H)= 7.11$\pm$0.10 by \cite{Kehrig2016}, but agree within uncertainties. For spaxels containing \oiv\ emission, we measure the global average three-state metallicity for \izw\ to be, 12+log(O/H)=7.20 with a dispersion of 0.08 dex around the mean. For the two-state and  metallicities in the same spaxels, we measure 12+log(O/H)=7.18 and 12+log(O/H)=7.16 with dispersion 0.08 dex and 0.10 dex, respectively. Again, the dispersion for the simple metallicities are larger w.r.t the two- and three-state metallicities.

Next we discuss the metallicities of the NW and SE star forming complexes. Measured inside the 1\arcsec\ radius aperture (white circle in Figure \ref{fig:FoV}), we find that the NW star forming complex has an average three-state metallicity of 12+log(O/H)$=$7.13$\pm$0.06 dex. The same region also exhibits the highest fraction of gas in the doubly-ionized helium state (see Figure \ref{fig:metal_maps}, right panel) indicative of a hard radiation field. Measured in the 1\arcsec\ radius aperture (yellow circle in Figure \ref{fig:FoV}), the SE complex exhibits an average three-state metallicity of 12+log(O/H)$=$7.2 $\pm$ 0.02 dex. Considering the smaller dispersion of the SE, we conclude that the metallicity of the SE star forming complex is homogenized with respect to the NW. The IFS study by \cite{Nakajima2024arXiv} report a 2D metallicity map of \izw\ that conflicts with the map presented in this work. \cite{Nakajima2024arXiv} report values of 12+log(O/H)$\sim$ 7.4 and 12+log(O/H)$\sim$ 7.0 for the NW and SE star-cluster respectively. These metallicities suggest that the NW star forming complex is more enriched than the SE star forming complex, and is opposite of the metallicities reported here. 

\begin{figure}
    \centering
    \includegraphics[width=\linewidth]{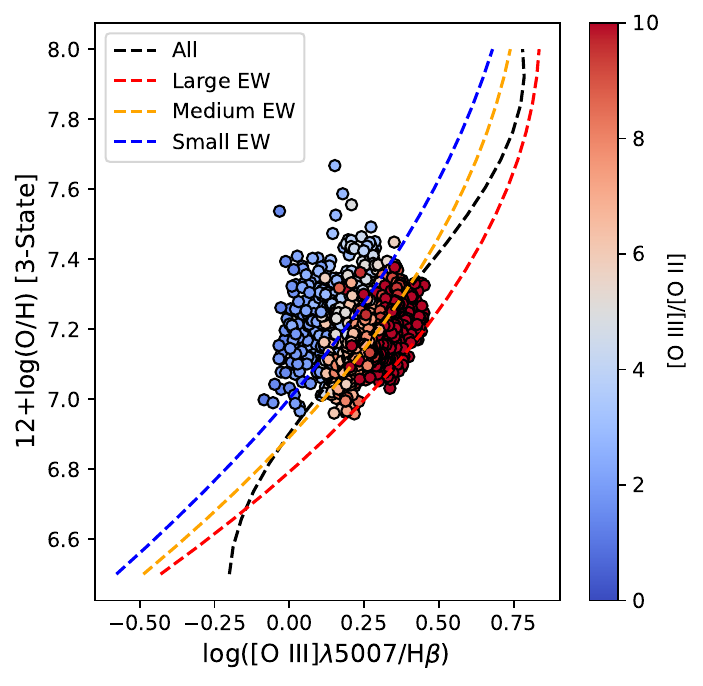}
    \caption{Strong-line calibrations against direct method metallicities. We show the three-state metallicities against the measured \oiii/\hb\ ratio. We overlay the \cite{Nakajima2022} \oiii/\hb\ calibrations for "Small" ($<$100 \AA, blue-dashed), "Medium" (100–200 \AA, orange-dashed), and "Large" ($>$200 \AA, red-dashed) H$\beta$ equivalent widths. Additionally we color-code the spaxel points by their ionization parameter as traced by their \oiii/\oii\ ratio. }
    \label{fig:Z_vs_R3}
\end{figure}

However, the metallicities from \cite{Nakajima2024arXiv} were indirectly inferred using a \oiii/\hb\ strong-line metallicity diagnostics calibrated from a combination of photoionization modeling and Sloan Digital Sky Survey single-slit optical spectroscopy of metal-poor galaxies \citep{Nakajima2022}. The ratio \oiii/\hb \ is also known to be correlated with ionization parameter \citep{BPT1981}. To account for the dual dependence of \oiii/\hb \ with metallicity and ionization parameter, \cite{Nakajima2022} use the equivalent width (EW) of H$\beta$ to correct for the ionization parameter dependence. To investigate how the \cite{Nakajima2022} calibrations compare to metallicities of \izw\ we show in Figure \ref{fig:Z_vs_R3} the three-state metallicities against the measured \oiii/\hb\ ratio. We also overlay the \cite{Nakajima2022} \oiii/\hb\ calibrations for ``Small EW" ($<$100 \AA), ``Medium EW" (100–200 \AA), and ``Large EW" ($>$200 \AA), and color-code the spaxel points by their ionization parameter as traced by their \oiii/\oii\ ratio. From this comparison, we find that the \cite{Nakajima2022} calibration only reproduces the observed metallicity for the highest ionization regions (i.e., \oiii/\oii $>$ 5). For regions with \oiii/\oii $<$ 5, the calibration underestimates the metallicity by up to 0.4 dex. Because the SE star forming complex exhibits a lower ionization parameter than the NW, the SE region will be more affected by this underestimate. This would explain why \cite{Nakajima2024arXiv}  report lower metallicities for the SE region than what is reported in this work. This comparison illustrates that spatially resolved metallicities derived from strong-line methods calibrated to integrated measurements can be uncertain. Instead, spatially resolved measurements should be calibrated on a spaxel-by-spaxel basis to achieve the highest accuracy possible. 

\begin{deluxetable*}{cccc}
\caption{Global and region average metallicities.}
\tablehead{\colhead{Region} & \colhead{12+log(O/H) $[$3-State$]$} & \colhead{12+log(O/H) $[$2-State$]$} & \colhead{12+log(O/H) $[$Simple$]$} \\ \colhead{} & \colhead{(dex)} & \colhead{(dex)} & \colhead{(dex)}}
\startdata
Global & 7.19 $\pm$ 0.13 & 7.18 $\pm$ 0.13 & 7.18 $\pm$ 0.15 \\
$[$O IV$]$ Spaxels\tablenotemark{a} & 7.20 $\pm$ 0.08 & 7.18 $\pm$ 0.08 & 7.16 $\pm$ 0.10 \\
NW Complex\tablenotemark{b} & 7.15 $\pm$ 0.06 & 7.08 $\pm$ 0.06 & 7.05 $\pm$ 0.07  \\
SE Complex\tablenotemark{c} & 7.22 $\pm$ 0.02 & 7.21 $\pm$ 0.03 & 7.18 $\pm$ 0.04 
\enddata
\tablenotetext{a}{Spaxels with \oiv$\lambda25.9$\micron\ emission S/N$>3$ }
\tablenotetext{b}{Measured inside a $r$=1\arcsec aperture centered on $\alpha=$ 09$^{\rm h}$~34$^{\rm m}_.$0~01$^{\rm s}_.$99, $\delta=$ 55$^{\circ}$~14\arcmin ~27.3\arcsec\ that is shown as the white circle in Figure \ref{fig:FoV}.}
\tablenotetext{c}{Measured inside a $r$=1\arcsec aperture centered on $\alpha=$ 09$^{\rm h}$~34$^{\rm m}_.$0~02$^{\rm s}_.$29, $\delta=$ 55$^{\circ}$~14\arcmin ~23.5\arcsec\ that is shown  as the yellow circle in Figure \ref{fig:FoV}.}
\label{tab:metal_table}
\end{deluxetable*}

\subsubsection{Chemical inhomogeneity within \izw}
\label{sec:chemical_inhomogeneity}
\begin{figure}
\centering
    \includegraphics[width=\linewidth]{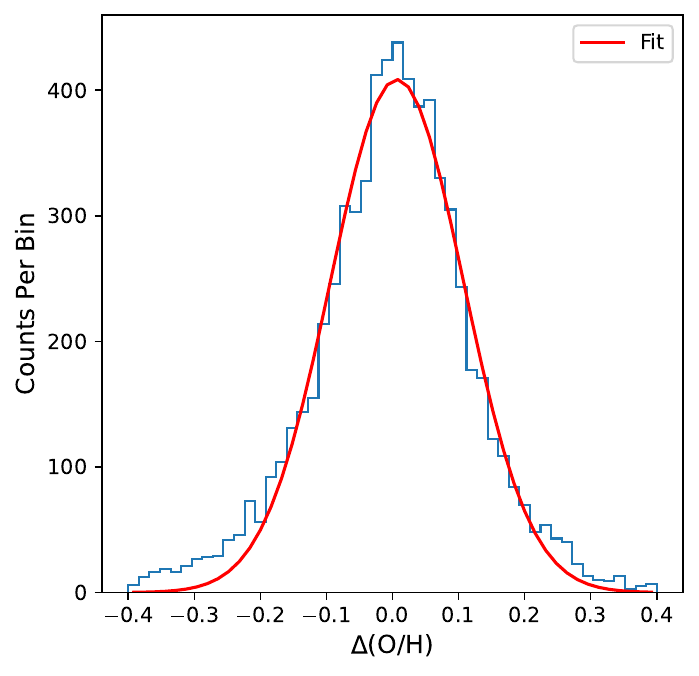}
    \caption{Distribution of the residual oxygen abundance after subtracting the mean three-state metallicity, 12+log(O/H)=7.19. The red-line is a Gaussian fit to the distribution.}
    \label{fig:metal_map}
\end{figure}

The metallicity map, shown in the left panel of Figure \ref{fig:metal_maps}, reveal pockets of metal-poor/rich gas in \izw\ and is suggestive of chemical inhomogeneities within the galaxy.
To quantify the degree of inhomogeneity in \izw's metallicity we follow criterion previously used on other dwarf galaxies including \izw\ \citep{PerezMontero2011,Kehrig2013,Kehrig2016}. First we check if the distribution of the residual three-state metallicities can be modeled by a Gaussian distribution. Shown in the right panel of Figure~\ref{fig:metal_map}, after subtracting the global mean metallicity, we construct the distribution of the metallicity residuals  ($\Delta$O/H) and fit a Gaussian model to the distribution. 
The Gaussian fit is best described with a width of $\sigma =0.13$. The best-fit $\sigma$ is equal to dispersion of the metallicity residuals. Despite the equality between $\sigma$ and the measured dispersion, we observe a tail of negative residuals skewed 3$\sigma$ from the global mean. 

Next, we perform a Lilliefors test \citep{Lillfors}. The null hypothesis of this statistical test is that the physical property in question can come from an unspecified Gaussian distribution. We perform this test using the \texttt{lilliefors} function within the Python module \texttt{statsmodel}\footnote{Available at \href{https://www.statsmodels.org/stable/index.html}{statsmodel}.}. The results of the Lilliefors test rejects the null hypothesis (i.e., that the abundance is homogeneous) with p-value $\simeq$ 10$^{-3}$. The low p-value suggest that the metallicity of \izw\ is not well described by a Gaussian distribution and that the galaxy exhibits chemical inhomogeneity across the observed field.

The result of the Lilliefors test disagrees with a previous IFS investigation into the presence of chemical inhomogeneity in \izw\ by \cite{Kehrig2016}. In their study, the authors found that the Lilliefors test was unable to reject the null hypothesis at a level of 36\%. However, this disagreement can be attributed to the different spatial sampling and/or spatial resolution \citep{Mast2014}, as the \cite{Kehrig2016} IFS data were obtained using the Potsdam Multi-Aperture Spectrophotometer (PMAS) with fibres sampling the galaxy at 1\arcsec\ scales. To test this, we re-bin the KCWI metallicity map to $\sim 1$\arcsec spatial sampling and convolve the map to 1\arcsec\ FWHM in order to mimic the best case scenario for PMAS observations \citep{Stanishev2012_alto_seeing,Sanchez2023}\footnote{\citet{Kehrig2016} do not report their observed seeing.}. After resampling, we repeat the Lilliefors test and arrive at the same conclusions as \cite{Kehrig2016}, finding that the test is unable to reject the null hypothesis at a level of 20\%.  This test demonstrates the power of high spatial resolution when deriving an accurate understanding of the metallicity distribution in galaxies.

To determine the physical scale of the inhomogeneities, we determine which spatial sampling, at 1\arcsec\ seeing, is required for the Lilliefors test to return a p-value $\sim$ 0.05 (or 5\%). We iterate between spatial samplings 60 pc (i.e., KCWI seeing) and 100 pc. Prior to resampling to metallicity map, we add to the metallicity map Gaussian distributed noise equal to the metallicity error. We find the spatial sampling at which the 5\% threshold is surpassed is already surpassed around $\sim$ 60 pc corresponding to the KCWI spatial resolution and the 86 pc resolution of MIRI/MRS, suggesting that the scale of inhomogeneities is at least 60 pc or smaller. We confirm that the above result is insensitive to the absence of O$^{3+}$/H$^+$ in the total abundance. For these reasons, we place an upper limit of 60 pc as the scale of inhomogeneities in \izw. 

\subsubsection{Are chemically enriched flows present in \izw?}
\label{sec:enirched_flows}
Due to the increased spatial resolution and field of view covered by KCWI, the metallicity map, shown in the left panel of Figure \ref{fig:metal_maps}, reveals for the first time metal-rich filamentary structure surrounding the star forming complexes of \izw. The filaments exhibit metallicities that vary between 12+log(O/H)$\sim 7.1-7.4$, and appear to connect to the SE and NW star forming complexes. The NW and SE filaments, labeled in Figure \ref{fig:metal_maps}, are tendrils of enriched gas that appear to connect to their respective star forming complexes. We also identify metal-enriched shells, with metallicities 12$+$log(O/H)$\gtrapprox$7.3 dex, southwest and northeast of the ionized gas shell traced by H$\beta$ emission as well as surrounding the metal-poor pockets. 

\begin{figure}
    \centering
    \includegraphics[width=\linewidth]{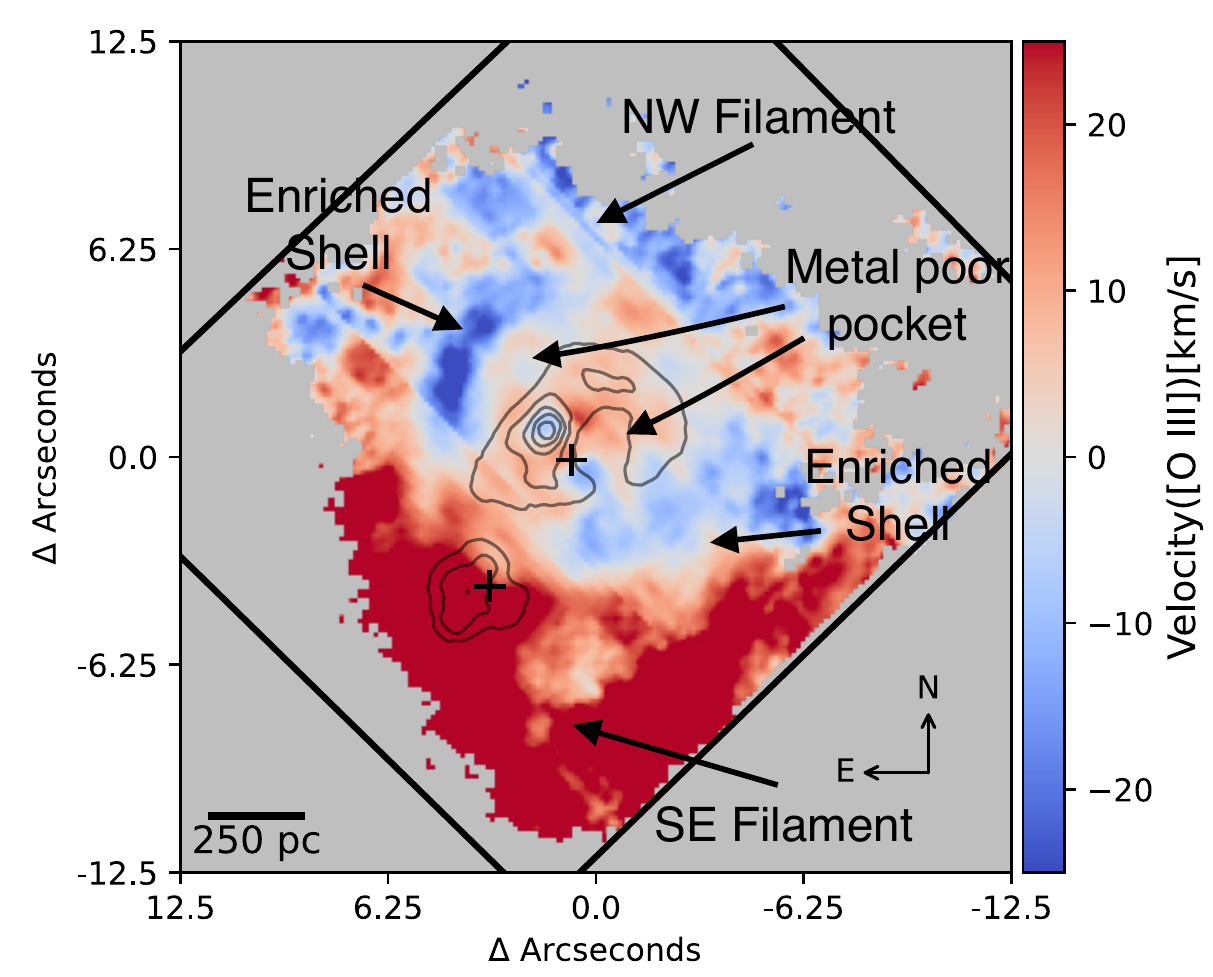}
    \caption{The radial velocity as measured from fits to the \oiii$\lambda$5007 emission. In addition to the H$\beta$ contour of the ionized gas shell, we overlay the positions of the SE/NW star forming complexes and the locations of metallicity inhomogeneities.}
    \label{fig:kinematic_map}
\end{figure}

Previous high velocity resolution observations of \izw\ provide evidence for the presence of a large scale outflow \citep{Lelli2012A,Arroyo-Polonio2024} and expanding shells \citep{Martin1996}. To explore whether or not such processes are producing the filaments or shells, we compare the metallicity map to the kinematics of the galaxy using the \oiii$\lambda5007$ velocity map shown in Figure \ref{fig:kinematic_map}. Generally, the velocity field revealed by the \oiii\ velocity map follow the \ion{H}{1} and H$\alpha$ kinematics mapped in \cite{Lelli2012A}, \cite{Petrosian1997}, and \cite{Arroyo-Polonio2024}. These studies mapped solid-body rotation where the bulk flow of gas is redshifted/blueshifted near the vicinity of the SE and NW star forming complexes. Like \cite{Arroyo-Polonio2024}, our maps show additional kinematic features  such as patches of redshifted ($\sim 10$ km s$^{-1}$) material superimposed on top of the blueshifted material associated with the bulk rotation. 

We are unable to disentangle the velocity field underlying the SE filament from the bulk rotation. The velocity field underlying the NW filament fluctuates between $\pm$ 5 km s$^{-1}$ around the systemic velocity \citep[$V_{\rm sys}\sim$751 km s$^{-1}$;][]{Thuan1999A&AS}. Such fluctuations are not indicative of a structured outflow/inflow. Alternatively, the fluctuations in the velocity field along may be tracing the expansion of compact \hii\ regions embedded within the  NW filament as proposed by \cite{Petrosian1997}. If this is the case, then the metal enhancements along the filament may be tracing the injection of metals from their star formation processes. Upon comparing the metal map to the kinematics map, we note that the enriched shells are co-spatial with the gas associated with the solid-body rotation. Where we observe one of the metal-poor pockets, and the doughnut structure of the ionized gas shell, we find that kinematics of the gas are redshifted up to 10 km s$^{-1}$ with respect to systemic velocity and at 20 km s$^{-1}$ from the bulk rotation. Because this gas is superimposed upon the bulk rotation, this redshifted gas may be evidence of feedback driven expansion arising from the NW star forming complex. But it is unclear if the feedback is actively removing gas and forming the metal-poor pockets since the ionized gas shell, which exhibits low dispersion in its metallicity, is also redshifted to similar velocities. The second metal-poor pocket is co-spatial with the blueshifted, bulk rotating gas, suggestive that this gas is connected to the extended structure of the galaxy and is in front or behind the NW star forming complex.

Generally, the magnitude of the redshifted/blueshifted velocities where we observe metallicity inhomogeneities are not high enough to escape the galaxy \citep{Martin1996}, suggesting that these are local features possibly driven by stellar feedback originating from the NW star-forming complex. Absent such feedback, the metals would likely be fully mixed on 100 pc scales because the age of the most recent burst of star formation \citep[$\sim$ 10 Myr;][]{Hirschauer2024,Bortolini2024} is older than the $\sim$ 1 Myr timescale for mixing via diffusion processes \citep{Roy1995}. 

Due to the observed X-ray source in the vicinity of the metal-poor pocket \citep[e.g.,][]{Kehrig_2021}, and the presence of a \ion{O}{8} hydrogen-like emission line in the X-ray spectrum \citep{Bomans2002ASPC..262..141B,Thuan2004ApJ...606..213T}, one alternative explanation for the metal-poor pocket is that an appreciable amount of oxygen may be locked in a hot-phase and unobservable in the optical. However, the highest temperature gas measured from the \oiii$\lambda$4363, see Appendix \ref{appendix:den_temp_maps} Figure \ref{fig:temp_map}, the peak of \oiv\ emission in Figure \ref{fig:very_high_maps}, and X-ray point source in Figure \ref{fig:FoV}, is offset from the location of the metal-poor pocket. Combined, this suggests that the metal-poor pocket is not co-spatial with the hottest and most highly excited gas where we may expect a hot-phase medium.

The smaller disturbances located in the outer regions of the galaxy may be associated with compact \hii\ regions. Despite the detection of an outflow in a UV-absorption line study of the galaxy \citep{Chisholm2016, Chisholm2018}, and expanding shells from echelle spectroscopy \citep{Martin1996}, we are not able to connect the enriched filaments or enriched shells to these dynamic processes as we are unable to separate their kinematics from the bulk rotation of the galaxy. 

\subsection{The effect of assumed gas properties on chemical abundances}
\label{sec:3_vs_2_zone_metallicity}
\begin{figure*}[t]
    \centering
    \includegraphics[scale=0.75]{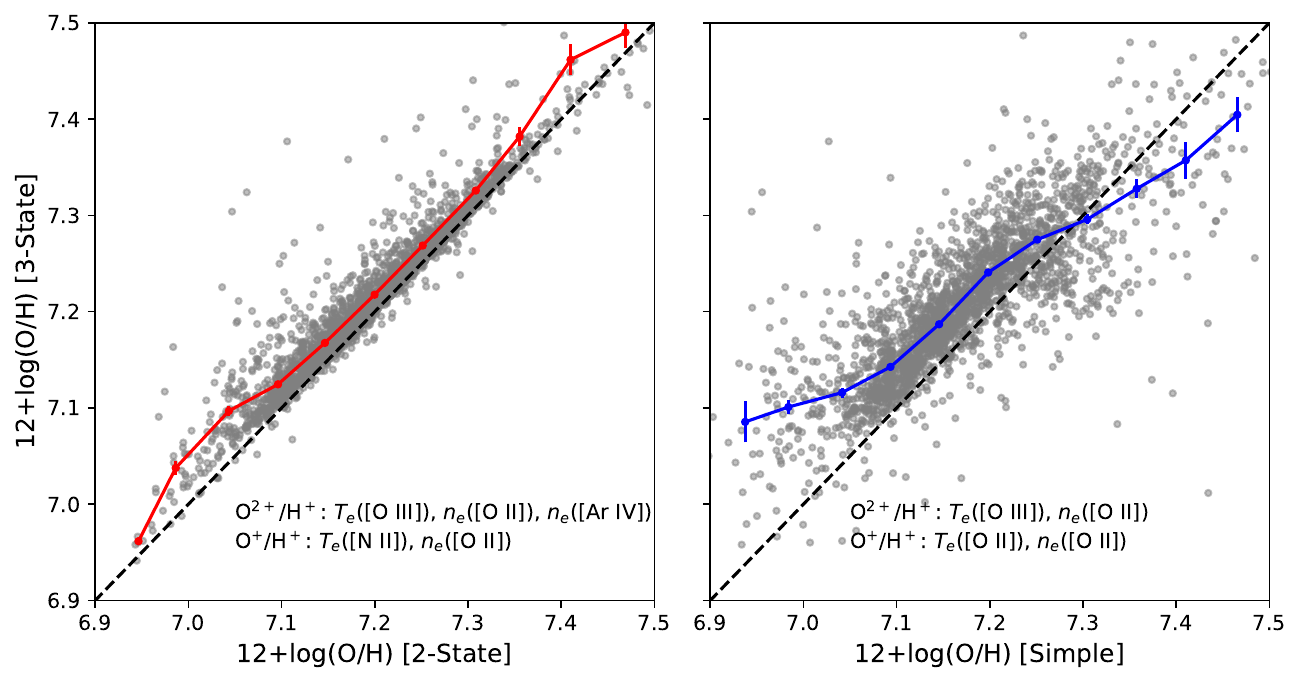}
    \caption{Investigating how different assumed properties impact derived metallicities. The three-state metallicities which include O$^{3+}$/H$^{+}$ are the same in each panel. Left: Pixel-to-pixel comparison of the three-state and two-state metallicities. The three- and two-state metallicities are derived using both low and high ionization gas densities as well as the temperature trend from \cite{Arellano-Cordova2020}. The red curve shows the average binned (0.05\,dex wide) metallicities. Right: Pixel-to-pixel comparison of the three-state and the simple metallicity derived assuming a single density diagnostic and the temperature trend from \cite{Campbell1986}. The blue-line are the binned average metallicities. In each panel, we annotate the properties used in the derivation of the two-state and simple metallicities. To aid the eye, we include a black-dashed line to indicate the one-to-one trend as well as annotate the assumed densities and temperatures for the singly and doubly ionized oxygen abundances. }
    \label{fig:z_vs_z}
\end{figure*}

We investigate how the three-state metallicities (i.e., the inclusion of O$^{3+}$/H$^{+}$) compares to the two-state and simple metallicities. In Figure \ref{fig:z_vs_z}, we show the pixel-to-pixel comparisons between the three-state, two-state, and simple metallicities. We discuss first the comparison between the three-state and two-state metallicity (derived using $n_{e}$([\ion{O}{2}]), $n_{e}$([\ion{Ar}{4}]), $T_{\rm e}$([\ion{N}{2}]), and \tempoiii). As shown in the left panel of Figure \ref{fig:z_vs_z}, the three-state metallicities largely lie above the one-to-one line. The offset between the two- and three-state metallicity, $\Delta{\rm O_{2,3}}$, appears to be mainly insensitive to the metallicity value. To view the average trend, we bin the data in 0.05 dex metallicity bins. The binned trend shows a slight increase toward higher three-state metallicities for two-state across the entire range of metallicities probed. The average difference between the two metallicities below this value is $\Delta{\rm O_{2,3}}=0.04\pm0.03$ dex. Above 12+log(O/H)=7.1 dex, the difference is $\Delta{\rm O_{2,3}}=0.02\pm0.03$ dex. Both differences are within their counterpart's uncertainty, suggesting that three-state metallicities are approximately 0.02-0.04 dex larger than two-state metallicities over the entire O/H range probed here. 

In the right panel of Figure \ref{fig:z_vs_z}, we compare the three-state metallicities to the simple metallicities (derived using $n_{e}$([\ion{O}{2}]), $T_{\rm e}$(\oii), and \tempoiii). We find here that the simple metallicities are systematically lower than the three-state metallicities. Unlike the two-state metallicity comparison, the difference between the three-state and simple metallicities becomes significant starting at 12+log(O/H)=7.3 dex, and increases for decreasing metallicities. The average difference between the two metallicities below 12+log(O/H)=7.1 dex is $\Delta{\rm O_{2,3}}=0.08\pm0.07$ dex. Above 12+log(O/H)=7.1 dex, the difference is $\Delta{\rm O_{2,3}}=0.04\pm0.05$ dex. These differences are $\sim$2 times larger than the differences involving the two-state metallicities whose nebular properties were carefully considered with respect to ionization. Like previous studies that derive ionic abundances using multi-ion density and temperatures appropriate for the ionization potential of each ion \citep[e.g.,][]{Berg2015ApJ...806...16B,Berg2020,Berg2021,Mendez-Delgado2023, Hayes2025}, these results show that two-state metallicities that assume the low-ionization density for high-ionization ions and/or rely on potentially biased temperature relationships can underestimate metallicities, particularly for 12$+\log$(O/H)$\,\leq\,7.3$.

\subsection{A recalibration of the oxygen ICF}
\label{sec:icfs}
We are in a unique position to compare the abundance of triply-ionized oxygen abundances to those predicted from theoretical ionization correction factors. Without observations of \oiv$\lambda25.89\mu$m emission, or other O$^{3+}$ transitions (e.g., \ion{O}{4}$\lambda$1407.38) the three-state oxygen abundance can be estimated using an ICF. ICFs are a factor that relates a subset of an element's ions, or transitions from species with similar ionization potential, to the element's total abundance. 
From photoionization models that sample metallicities down to \izw's \citep[$Z$=\text{[}0.2, 0.05, 0.02\text{]}$\times Z_{\odot}$;][]{Stasinka2003}, \cite{Izotov2006} constructed the following ICF to estimate the O$^{3+}$/H$^{+}$ abundance using the fractional abundance of doubly ionized helium:
\begin{equation}
\label{eqt:ICF}
    \frac{\rm O^{3+}}{\rm H^{+}} =\alpha \left(\frac{\rm He^{2+}}{\rm He^{+}+\rm He^{2+}}\right)\left( \frac{\rm O^{+}+\rm O^{2+}}{\rm H^{+}} \right),
\end{equation}
where $\alpha=0.5$. 

\begin{figure}
    \centering
    \includegraphics[width=\linewidth]{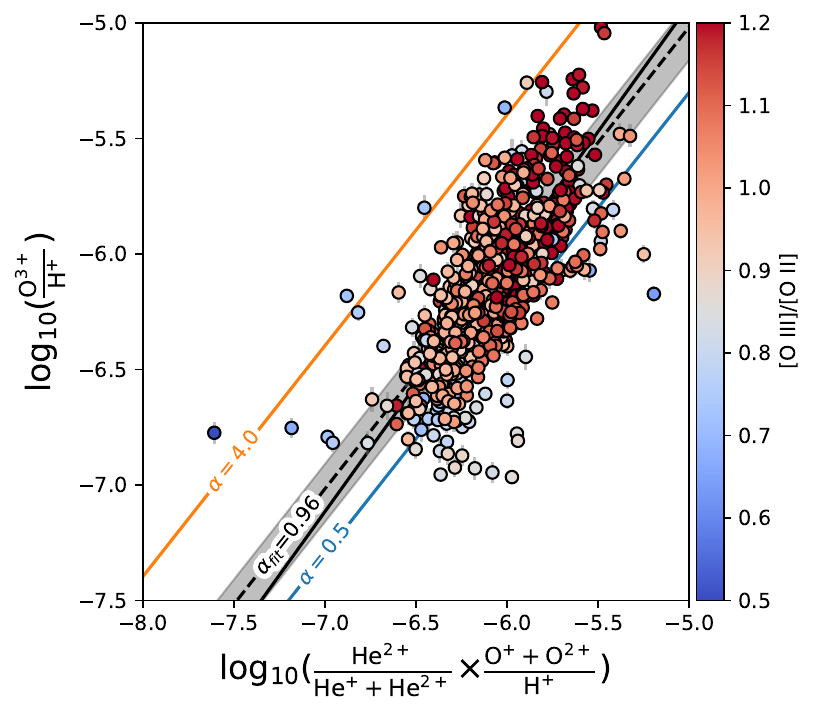}
    \caption{The two-state oxygen ionization correction factor against the triply-ionized oxygen abundance. The data points are colored by level of ionization traced by the \oiii/\oii\ ratio. We the overlay theoretical ionization correction factor \cite{Izotov2006} with $\alpha=0.5$ as well as $\alpha=4$. We also show a black-dashed line indicating the best-fit ICF, $\alpha=0.96\pm0.1$. The shaded grey is the $\pm3\sigma$ of the fit. The solid black line is the fit if we allow a non-unity power-law index.}
    \label{fig:ICF}
\end{figure}

To test if the oxygen ICF, see Eq. \ref{eqt:ICF}, predicts the observed O$^{3+}$ abundance we compare the measured O$^{3+}$/H$^{+}$ against abundances predicted using Eq. \ref{eqt:ICF} in Figure \ref{fig:ICF}. In addition to color-coding the markers by \oiii/\oii, we overlay the ICF of \cite{Izotov2006} as well as the ICF that reproduces the highest values of O$^{3+}$/H$^{+}$. The $\alpha$ scaling that contains the observed triply-ionized oxygen abundances ranges between 0.5 and 4.0. We determine the scaling factor, $\alpha_{\mathrm{fit}}$, which minimizes the residuals between the triple-ionized oxygen abundance and the ICF. We find the best fit scaling to be $\alpha_\mathrm{fit}=0.96\pm0.1$. This value is almost a factor of two larger than the value determined from the stellar photoionization models reported in \cite{Izotov2006}, and implies that previous determinations of O$^{3+}$/H$^{+}$ using this ICF may be underestimated by $\sim$0.5 dex, particularly for objects with very high ionization parameter. 

There is evidence that relationship between log$_{10}$(O$^{3+}$/H$^{+}$) and the ICF may require a power-law index, $n$, of non-unity. When we re-fit the line, and allow the power law index to vary, we find $\alpha_\mathrm{fit}=3.64\pm0.03$ and power-law index, $n=1.09\pm0.01$. The power law is given by the solid black line in Figure \ref{fig:ICF}. Additionally, within Figure \ref{fig:ICF}, we observe that there is large dispersion in O$^{3+}$/H$^{+}$, $\sigma_{\rm O^{3+}}$, for fixed values of the ICF. We compared the dispersion around the trend line with ionization parameter as traced by  \oiii/\oii\ and the hardness of the radiation field as traced by \heii/\hb. We find neither of these parameters is correlated with the dispersion. 
A similar discrepancy between the observed and predicted triply-ionized oxygen abundance has been reported by \cite{Berg2021} for two galaxies with similar mass and metallicity as \izw. They measured O$^{3+}$/H$^{+}$ using \ion{O}{4}$\lambda$1407.38 emission measured in the integrated UV spectrum of the galaxies J104457 and J14185. To compare the observed abundances to theoretical expectations, they ran a \texttt{CLOUDY} version 17.00 \citep{Ferland2013RMxAA..49..137F} photoionization models covering ionization parameters, $-3<\log\mathcal{U}<-1$, metallicity $7.4<$ 12+log(O/H) $<8.4$, and densities $n_{\rm{e}}=10^1-10^4$ cm$^{-3}$, using Binary Population and Spectral Synthesis \citep[BPASSv2.14;][]{Eldrige2016Bpass, Stanway2016Bpass} templates as ionizing sources. As the metric for comparison, \citet{Berg2021} calculate the O$^{3+}$/O$^{2+}$ abundance ratio for each galaxy, 0.04 and 0.02 for J104457 and J14185 respectively. When compared to the model abundances, they found that models underestimate the observed O$^{3+}$/O$^{2+}$ ratio by more than an order of magnitude over metallicities and ionization parameter between 0.05-0.30$Z_{\odot}$ and $-4< \log \mathcal{U} <-1$. To place \izw\ in the context of those results, we compute the average O$^{3+}$/O$^{2+}$ ratio and find O$^{3+}$/O$^{2+}=0.07\pm0.03$. This ratio agrees with the ratio measured for J104457 and J14185, and implies that \izw's abundance of triply-ionized oxygen would not be reproduced by models presented in \cite{Berg2021}. 

A central assumption in the evidence presented for setting $\alpha\sim1.0$ in the oxygen ICF is that the entirety of ionizing photons producing O$^{3+}$ originate from stellar sources and not additional sources such as fast radiative shocks, Pop III stars, Active Galactic Nuclei (AGN) powered by an Intermediate Mass Black Hole (IMBH). Despite exhaustive searches \citep[e.g.,][]{Stasinka1999A&A...351...72S,Thuan2005ApJS..161..240T,Kehrig2015ApJ...801L..28K, RickardsVaught2021}, identification of such sources to explain the very high-ionization emission remain elusive. The discrepancy between observed high-ionization abundance and predictions from theoretical ICFs based standard stellar populations illustrate the need to improve our understanding of the physics controlling the high energy ionizing flux output by standard stellar populations in low-metallicity environments.

\section{Discussion}
\label{disc:main}
\subsection{On the chemical inhomogeneities observed in \izw}
Modern IFS observations of dwarf galaxies are inconclusive on the pervasiveness of chemical inhomogeneities in dwarf galaxies. Many studies have suggested that their chemical abundances are homogeneous on physical scales $>$ 100 pc \citep{James2010,PerezMontero2011,Kehrig2013,Kumari2019}, including \izw\ \citep{Kehrig2016}.  Other studies find significant inhomogeneities \citep{James2013,James2020,Cameron2021_DUVET,McPherson2023_DUVET,Peng2023} at similar physical scales. In the case of \izw, our analysis finds that \izw\ is inhomogeneous at the physical scales probed by KCWI ($\sim$60 pc) and that observations of \izw\ by \cite{Kehrig2016} lacked the spatial resolution to observe inhomogeneities at these scales. The conflicting results suggest that while IFS capabilities may have advanced the ability to measure inhomogeneities for large physical scales, even finer resolution is required to capture small scale or isolated disturbances in dwarf galaxies \citep[see][]{Mast2014,James2020}.

The source(s) of oxygen abundance inhomogeneities in dwarf galaxies have been tied to several astrophysical processes including: outflows \citep{James2013,McQuinn2020,James2020,Cameron2021_DUVET,McPherson2023_DUVET,Peng2023}, accretion of metal-poor gas \citep{delValle-Espinosa2023,Nakajima2024arXiv}, and mergers \citep{Pascale2022}, but in some cases the source of metal-enrichment/mixing is unclear \citep{Cosens2024}. Of the listed mechanisms, outflows/inflows are the favored mechanism to explain why numerous galaxies, including \izw, lie below the mass-metallicity relationship \citep[see,][]{Hirschauer2016ApJ...822..108H,Chisholm2018,McQuinn2020}. Despite inferred presence of outflows via UV absorption-lines \citep{Chisholm2018}, and \ion{H}{1} gas kinematics \citep{Lelli2012A}, we are unable to directly connect the observed inhomogeneities in \izw\ to either outflow or inflow features. Given the lack of large-scale kinematic disturbances that would be indicative of outflows/inflows, one plausible scenario is that stellar feedback transported metal-rich gas away from the galaxy's star forming regions \citep[e.g.,][]{Dale2005}. Because the relative abundances of other ions with oxygen (e.g., N/O and S/O) better trace the contribution of different stellar processes to the overall metal mixing \citep{Tenorio-Tagle1996,Krumholz2018, Emerick2020} a more comprehensive analysis of the sources disturbing the galaxy's metals requires matched resolutions maps of  N and/or S abundances. An analysis of this type would not only improve our understanding of feedback processes in low metallicity environments, but would also increase our understanding of metal-mixing processes currently unresolved in high-$z$ galaxies. 

Because \izw\ is one of the nearest metal-poor galaxies, it represents the best-case scenario with existing facilities for achieving the requisite spatial resolution to measure inhomogeneities. These results therefore stress the need for improving the technical ability to resolve dwarf and high-$z$ galaxies in order to better characterize the magnitude and scales of chemical inhomogeneities in order to better understand the processes driving galaxy chemical evolution.

\subsection{Uncertainties in oxygen abundances calculated due to assumed properties or strong-line methods}
Observational challenges (i.e., faint emission, narrow wavelength coverage) in obtaining the multi-ionization zone diagnostics required for an accurate characterization of the chemical abundances within high-$z$ galaxies means that investigators \citep[e.g.,][]{Sanders2024,Scholte2025arXiv} sometimes have to rely on a single density and/or temperature relations that can be biased due to shocks, as well as unresolved temperature and/or density inhomogeneities \citep{Binette2012,Mendez-Delgado2023,Mendez-Delgado2023Nature, RickardsVaught2024}, to derive metallicities. Here we derive, and compare, metallicities using a combination of different assumptions for the physical properties of the gas. The different assumptions (i.e., three-state, two-state, and simple metallicities) are detailed in Section \ref{meth:oxygen_abundance}. As shown in the global and region-by-region metallicities tabulated in Table \ref{tab:metal_table}, on global scales, the set of three metallicities are indistinguishable within the 0.1 dex uncertainty. However, on resolved and spaxel-by-spaxel scales, the ``simple" metallicities not only exhibit the largest scatter, but also underestimate the three-state metallicity by $\sim 0.1$ dex. 

The accuracy of high-$z$ metallicities becomes increasingly uncertain when strong-line calibrations are the only available metric to derive metallicity. The \oiii/\hb\ ratio (i.e., $R_3$), and other calibrations, are often employed in high-$z$ studies \citep{Sarkar2025_mzr_calibrations, Curti2024, Boyett2024,Bisigello2025} to investigate properties of galaxy evolution (e.g., mass-metallicity relationship). In Section \ref{res:global_metals} Figure \ref{fig:Z_vs_R3}, we compared, on a spaxel-by-spaxel basis, the three-state oxygen abundances to the \cite{Nakajima2022}  $R_3$ calibration used to infer \izw's metallicity in \cite{Nakajima2024arXiv}. We found that the $R_3$ calibration under-estimates the metallicity for regions with lower-ionization parameter as traced by \oiii/\oii. In spite of considerable effort towards calibrating high-$z$ direct-method abundances to calibrations \citep{Sanders2024, Laseter2024}, the disagreement between direct and calibrated metallicities suggests caution when interpreting abundances based on strong-line methods.

\subsection{Uncertainty in high-ionization ICFs}
\label{dis:high_ion}
Identifying the ionizing source(s) producing very high-ionization species has been a long standing problem for \izw\ \citep{Izotov1998ApJ...497..227I,Kehrig2016, RickardsVaught2021} and other BCDs \citep[e.g.,][]{Thuan2005ApJS..161..240T,Kehrig2018,Wofford2021}. While many of these studies largely attempted to explain intense He\,\textsc{ii} emission in these galaxies, more recent observations are finding unexplained enhancements of \oiv\ \citep{Berg2021, Olivier2022} and even higher ionization potential ions such as Ne$^{4+}$ \citep[][Hunt et al. in prep]{Mingozzi2025arXiv}. Known colloquially as the high-energy ionizing photon production problem \citep[HEIP$^3$;][]{Berg2021}, the observed discrepancies with modeling suggests that either the high energy ionizing flux of stellar populations is unknown, or sources beyond stars are needed to reproduce the abundance of high ionization species \citep{Leitherer1999,Feltre2016}.

Sources that have been considered as producers of high-ionization photons include: fast radiative shocks \citep{Thuan2005ApJS..161..240T}, Wolf-Rayet stars \citep{Naze2003A&A...401L..13N,Naze2003A&A...408..171N}, Ultraluminous X-ray sources \citep{Moon2011ApJ...731L..32M,RickardsVaught2021,Simmonds2021,Garofali2024}, X-ray binaries \citep{Schaerer2019A&A...622L..10S}, theoretical Population III stars \citep{Kehrig2016}, and IMBH \citep{richardson2022}. However, modeling of these objects are uncertain and generally reproduce observations if the object is the sole ionizing source \citep{Gutierrez2014}, or under specific parameter space \citep[WR stars with clumpy and optically thin winds;][]{Roy2025arXiv}. Even in cases where models of exotic sources \citep{Allen2008ApJS..178...20A, Feltre2016, richardson2022, martinezparedes2023, flury2024} can explain the most extreme ionization emission (e.g., [\ion{Ne}{5}]), the source models show other limitations such as failing to reproduce line emission from lower ionization species \citep[see,][Paper\,I]{Mingozzi2025arXiv}.  Our recalibration of the oxygen ICF adds to the growing body of evidence that widely used stellar models alone do not generate the requisite number of high-ionization photons $> 54$ eV needed produce the abundance of high-ionization species. Towards increasing our understanding of the source(s) driving the production of such species, we will present a joint optical and mid-IR study of the ionizing structure and source within \izw\ in an upcoming analysis (Rickards Vaught et al, in prep).

\section{Conclusions}
\label{sec:conlusion}
We presented KCWI and MIRI/MRS observations of the very low metallicity galaxy \izw. As this galaxy is the quintessential high-$z$ analog, observations of \izw\ allow us to observe, at resolved scales, the physical properties, and ionizing conditions, of the low-metallicity ISM present in the early universe. Specifically, the joint KCWI and MIRI/MRS observations provide unique access to optical and mid-IR oxygen emission (e.g., \oii, \oiii, and \oiv), as well emission from other ions, which allow for robust measurements of electron temperature and multi-ionization zone densities. Using these derived properties, we constructed a tens of parsec scale map of the direct, three ionization state, oxygen abundance. From these observations our findings are as follows:
\begin{itemize}

    \item As discussed in Section \ref{sec:derivation_of_nebular_properties}, using nebular \oii$\lambda\lambda3726$, $3728$\AA, \ariv$\lambda\lambda4711$, $4741$\AA\ and \oiii$\lambda4363$\AA\ emission, we derived the electron density of the low-ionization zone, as well as the electron density and temperature of the high-ionization zone. Using the suite of oxygen emission lines covered within both the optical and mid-IR spectra as well as the derived physical gas properties, in Section \ref{meth:oxygen_abundance}, we presented a tens of pc scale resolved map of \izw's three-state oxygen abundance (i.e., including O$^{+}$/H$^+$, O$^{2+}$/H$^+$, and O$^{3+}$/H$^+$).

    \item In Section \ref{res:neb_emission}, we presented maps that show emission from low- and high-ionization ions extending out to distances $>$ 500 pc from \izw's NW and SE star forming complexes. From the joint MIRI/MRS and KCWI observations, we also presented maps of \oiv\ and \heii\ emission. The emission from both of these very high- ionization ions ($\sim$ 54 eV) is co-spatial and found in both the SE and NW star forming complexes.

    \item Presented in Section \ref{sec:helium_abundances}, we derived the abundances of singly and doubly ionized helium. We measured the helium mass fraction and found it to be consistent with previous measurements of the galaxy \citep{Izotov1998ApJ...497..227I}. Compared with previous determinations of the primordial helium abundance, the helium mass fractions are lower but consistent within uncertainties. We also presented a map of the fraction of doubly-ionized to total helium in Figure \ref{fig:metal_maps}. The fractional abundance is $>$10\% in regions inside the ionized gas shell and for a region with an undetected stellar population \citep[see][]{Hunter1995ApJ...452..238H, RickardsVaught2021}.

    \item  As discussed in Section \ref{sec:enirched_flows}, due to the finer spatial resolution afforded by KCWI, relative to previous IFS observations, we detect the presence of 60 pc scale chemical inhomogeneities within \izw. The inhomogeneities are characterized by pockets of metal-poor gas in and around the NW star forming complex, as well as shells/filaments of metal-enriched gas.
    
    \item In order to compare how different assumptions on the gas physical conditions impact inferred metallicities, we also derived ``two-state" (i.e., including only O$^{+}$/H$^+$, O$^{2+}$/H$^+$, and \tempoii=\tempnii) as well as ``simple" (i.e., single density and \tempoii\ from \cite{Campbell1986})  metallicities. In Section \ref{sec:3_vs_2_zone_metallicity}, we compared this set of metallicities to the three-state metallicity and found good agreement between the three-state and two-state metallicities. However, we found that ``simple" metallicities (derived using similar assumption to high-$z$ studies) underestimate the three-state metallicities by up to 0.1 dex, particularly for metallicities $<$ 12+log(O/H) $\sim$ 7.1 dex.
    
    \item In Section \ref{sec:icfs} we compared the observed O$^{3+}$/H$^+$ to those predicted from the oxygen ICF \citep{Izotov2006}. We found that \izw's  O$^{3+}$/H$^+$ abundance is underestimated by a factor of 2 when reproduced using the ICF constructed from simple stellar population photoionization models. This discrepancy implies that either the high ionizing flux of current simple stellar population models need to be improved, or other undetected ionizing sources (e.g., a low-luminosity AGN powered by an IMBH) are needed.
\end{itemize}

This joint MIRI/MRS and KCWI analysis of \izw\ demonstrates the importance of high spatial resolution observations of dwarf galaxies in order to uncover their ``true'' chemical abundance distribution. The extension to mid-IR wavelengths afforded by MIRI/MRS allowed access to high ionization species, which we showed can have non-negligible contribution to the total ionic abundance, and emphasize that optical spectral alone may not be enough to accurately assess the chemical abundance of galaxies. Future observations covering redder wavelengths will be important to better diagnose ISM gas properties as well as judge nitrogen and sulfur abundances relative to oxygen, important to diagnosing sources of chemical enrichment and metal gas flows. 
    
\begin{acknowledgments}
The authors thank the referee for their report that improved the clarity and analysis of the work presented here.
This work is based on observations made with the NASA/ESA/CSA James Webb Space Telescope. The data were obtained from the Mikulski Archive for Space Telescopes (MAST,\dataset[~doi: 10.17909/n80x-b534]{https://doi.org/10.17909/n80x-b534 }) at the Space Telescope Science Institute, which is operated by the Association of Universities for Research in Astronomy, Inc., under NASA contract NAS 5-03127 for JWST. These observations are associated with program JWST-3533. RRV is grateful for the support of this program, provided by NASA through a grant from the Space Telescope Science Institute.
The data presented herein were also obtained at the W. M. Keck Observatory, which is operated as a scientific partnership among the California Institute of Technology, the University of California and the National Aeronautics and Space Administration. The Observatory was made possible by the generous financial support of the W. M. Keck Foundation. The authors wish to recognize and acknowledge the very significant cultural role and reverence that the summit of Maunakea has always had within the indigenous Hawaiian community. We are most fortunate to have the opportunity to conduct observations from this mountain. We also wish to thank all the Keck Observatory staff for their observational support. 
BLJ and MM are thankful for support from the European Space Agency (ESA).  MM acknowledges that a portion of their research was carried out at the Jet Propulsion Laboratory, California Institute of Technology, under a contract with the National Aeronautics and Space Administration (80NM0018D0004). 

This work made use of Astropy:\footnote{http://www.astropy.org} a community-developed core Python package and an ecosystem of tools and resources for astronomy \citep{astropy:2013, astropy:2018, astropy:2022}. Additionally, this research made use of wcsaxes, an open-source plotting library for Python hosted at https://wcsaxes.readthedocs.io/en/latest/, matplotlib, a Python library for publication quality graphics \citep{Hunter:2007}, the IPython package \citep{PER-GRA:2007}, and NumPy \citep{harris2020array}. This research made use of ds9, a tool for data visualization supported by the Chandra X-ray Science Center (CXC) and the High Energy Astrophysics Science Archive Center (HEASARC) with support from the JWST Mission office at the Space Telescope Science Institute for 3D visualization. 
\end{acknowledgments}
\facilities{JWST (MIRI/MRS), Keck:II (KCWI)}
\software{wcsaxes, IPython \citep{PER-GRA:2007}, Astropy \citep{astropy:2013, astropy:2018, astropy:2022}, matplotlib \citep{Hunter:2007}, NumPy \citep{harris2020array}, ds9}

\appendix
\section{Example Fits to Blended Argon Emission Lines}
\label{Appendix:Blended_Argon}
\begin{figure}
    \centering
    \includegraphics[width=0.5\linewidth]{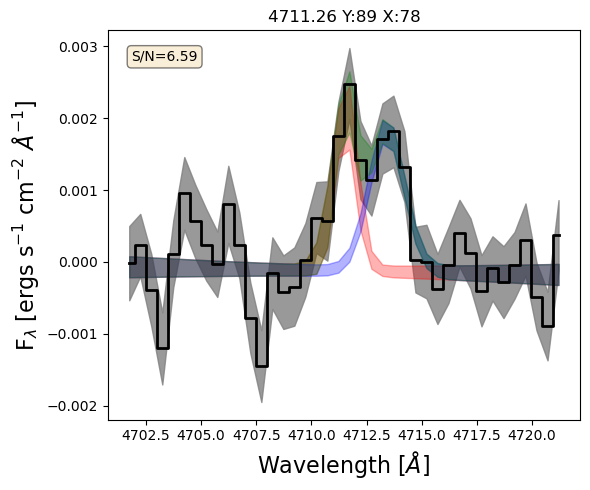}
    \caption{Example fit of the blended [Ar IV]$\lambda4711$ and He I $\lambda4715$ emission. The shaded red(purple) regions is the $1\sigma$ model uncertainty of the [Ar IV]$\lambda4711$(He I $\lambda4715$) fit, while the shaded green is the uncertainty on the total fit. The shaded grey is the uncertainty measured for the continuum. The [Ar IV]$\lambda4711$ emission in the KCWI pixel (89,78), within the NW star forming complex, is detected at S/N=6.59.}
    \label{fig:blended_ar_fit}
\end{figure}

The blended emission lines [Ar IV]$\lambda4711$/He I $\lambda4715$ are marginally resolved at the spectral resolution of KCWI. To fit their emission we perform a double-Gaussian plus linear continuum fit. In the particular case of [Ar IV]$\lambda4711$/He I $\lambda4715$ blend, to accurately remove the He I $\lambda4715$ contamination from [Ar IV]$\lambda4711$, we first fit He I $\lambda4471$ and assume a theoretical ratio He I $\lambda4471$/$\lambda4715=0.150$ ($n_{\rm e}=10^2$ cm$^{-3}$, $T_{\rm e}=20,000$ K), to fix the amplitude, velocity and width of the Gaussian fit to He I $\lambda4715$. Discussed in detail in Section \ref{meth:dust}, the dust extinction in \izw\ is low, so that for deblending purposes the observed ratios between different helium recombination lines should be close to their theoretical value. We show an example fit to the blended emission as well as the final [Ar IV]$\lambda4711$ emission-line map in Figure \ref{fig:blended_ar_fit}.

\section{Maps of Densities and High-Ionization Zone Temperature}
\label{appendix:den_temp_maps}
\begin{figure*}
    \centering
    \includegraphics[width=0.45\linewidth]{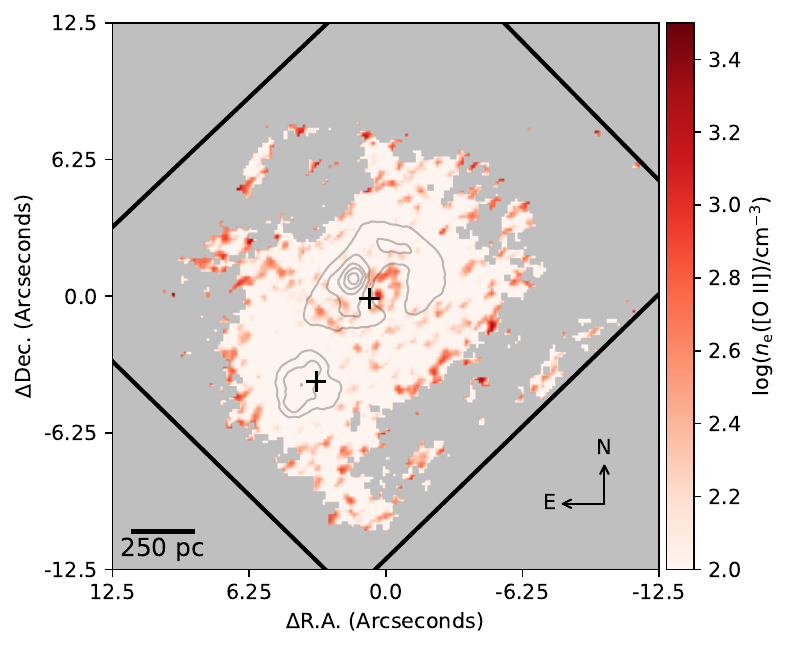}
    \includegraphics[width=0.45\linewidth]{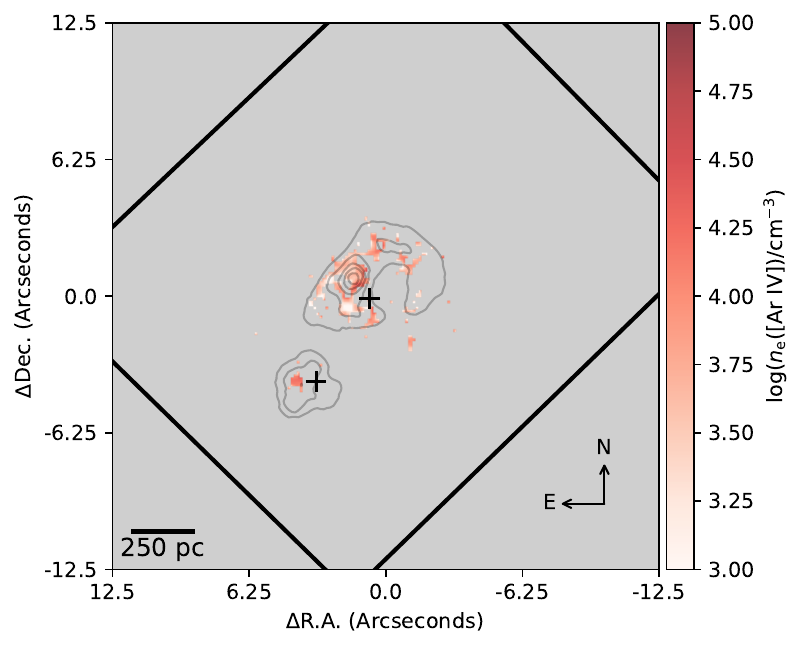}
    \includegraphics[width=0.37\linewidth]{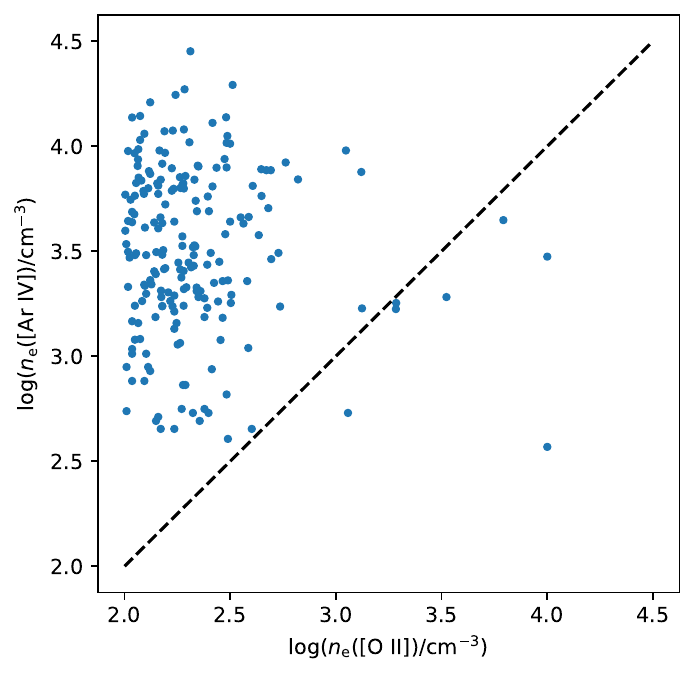}
    \caption{Maps and comparisons between the  low- and high-ionization zone densities. Top left: Map of low-ionization zone electron density as measured from the \oii$\lambda\lambda3726,3728$ doublet. Top right: Map of high-ionization zone electron density as measured from the \ariv$\lambda\lambda4711,4741$ doublet. Bottom: Spaxel-to-spaxel comparison between the  low- and high-ionization zone densities }
    \label{fig:density_maps}
\end{figure*}

Discussed in Section \ref{meth:electron_dens}, we derived multi-ion electron densities using the \oii$\lambda\lambda3726,3728$ and \ariv$\lambda\lambda4711,4741$ doublet. In Figure \ref{fig:density_maps}, we show the maps of the two inferred densities, as well as a one-to-one comparison. The \oii$\lambda\lambda3726,3728$ doublet is detected across a large area of the galaxy, and enables an estimate of the low-ionization zone density in the majority of spaxels. In contrast, the \ariv$\lambda\lambda4711,4741$ emission lines are detected only detected with sufficient significance for spaxels coincident with the SE star forming complex, as well as the ionized gas shell surrounding the NW star forming complex. As shown in the one-to-one plot, where both densities can be measured, the high-ionization zone electron density is systematically larger than the low-ionization zone density.

\begin{figure}
    \centering
     \includegraphics[width=0.5\linewidth]{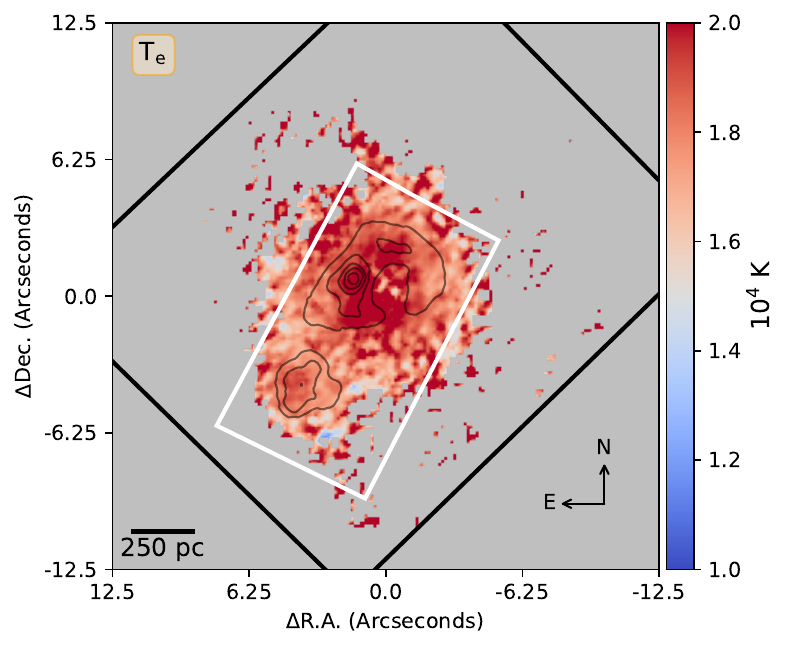}
    \caption{ Map of $T_{\rm e}$(\oiii) derived from measurements of the \oiii$\lambda$4363/$\lambda$5007 ratio. The R.A. and dec. offset are centered on the R.A. and dec. coordinates 09$^{\rm h}$~34$^{\rm m}_.$0~01$^{\rm s}_.$92, 55$^{\circ}$~14\arcmin ~27.4\arcsec.}
    \label{fig:temp_map}
\end{figure}

In Section \ref{meth:electron_temp}, we derived \tempoiii\ using the \oiii$\lambda4363/5007$ emission line ratio. We show in Figure \ref{fig:temp_map} the spatial distribution of the derived temperature. Generally, the temperature distribution appears to be homogeneous between 1.6 and 1.8 $\times 10^4$ K. However, the temperature is as high as 2.0$\times 10^4$ K for a region near the cavity of the ionized gas shell.

\bibliography{main}{}
\bibliographystyle{aasjournal}
\end{document}